\documentclass[useAMS,usenatbib]{mn2e}

\usepackage{float}
\usepackage{graphicx}
\usepackage{amssymb}
\usepackage{amsmath}
\usepackage{datetime}
\usepackage[normalem]{ulem}
\usepackage{times}
\usepackage[export]{adjustbox}

\usepackage{color}
\usepackage{soul}
\definecolor{stan}{rgb}{0,0,1}

\definecolor{eliot}{rgb}{0.7,0,1}

\newcommand{\apjs}{ApJS}
\newcommand{\apjl}{ApJ}

\newcommand{\aap}{A\&A}

\def\<<{{\ll}}
\def\>>{{\gg}}

\def\spose#1{\hbox to 0pt{#1\hss}}
\def\ltwig{\mathrel{\spose{\lower 3pt\hbox{$\mathchar"218$}}
     \raise 2.0pt\hbox{$\mathchar"13C$}}}
\def\gtwig{\mathrel{\spose{\lower 3pt\hbox{$\mathchar"218$}}
     \raise 2.0pt\hbox{$\mathchar"13E$}}}

\newcommand{\beq}{\begin{equation}}
\newcommand{\eeq}{\end{equation}}
\newcommand{\beqa}{\begin{eqnarray}}
\newcommand{\eeqa}{\end{eqnarray}}

\title[Super-Eddington winds]{
Super-Eddington stellar winds: unifying radiative-enthalpy vs.\  flux-driven models
}
\author[Owocki, Townsend \& Quataert]{Stanley P.\ Owocki$^{1,4}$,
Richard H. D. Townsend$^{2,4}$ and Eliot Quataert$^{3,4}$
\\
$^1$Bartol Research Insitute, 
Department of Physics \& Astronomy, 
University of Delaware, Newark, DE 19716, USA
\\
$^2$Department of Astronomy, University of Wisconsin, 5534 Sterling Hall
Madison WI 53706, USA
\\
$^3$
Astronomy and Physics Departments and Theoretical Astrophysics Center, University of California, Berkeley, Berkeley, CA 94720, USA
\\
$^4$
Kavli Institute of Theoretical Physics, University of California, Santa Barbara, Santa Barbara, CA 93106, USA
}

\begin{document}

\include{aas_macros}

%\date{Accepted ?.  Received ?; in original form ?}

%\pagerange{\pageref{firstpage}--\pageref{lastpage}} \pubyear{2011}

\maketitle

\label{firstpage}

\begin{abstract}
We derive semi-analytic solutions for optically thick, super-Eddington stellar winds, induced by an assumed steady energy addition $\Delta {\dot E}$ concentrated around a near-surface heating radius $R$
 in a massive star of central luminosity $L_\ast$.
We show that obtaining steady wind solutions requires both that the resulting total luminosity $L_{\rm o} = L_\ast + \Delta {\dot E}$ exceed the Eddington luminosity,
 $\Gamma_{\rm o}  \equiv L_{\rm o}/L_{\rm Edd} > 1$, 
and that the induced mass loss rate be such that the ``photon-tiring" parameter $m \equiv {\dot M} GM/R L_{\rm o} \le 1-1/\Gamma_{\rm o}$, ensuring the luminosity is sufficient to overcome the gravitational potential $GM/R$.
Our analysis unifies previous super-Eddington wind models that either: (1) assumed  a direct radiative flux-driving without accounting for the
advection of radiative enthalpy that can become important in such an optically thick flow; or (2) assumed that 
such super-Eddington outflows are adiabatic, neglecting the effects of the diffusive radiative flux.
We show that these distinct models become applicable in the asymptotic limits of small vs.\ large values of $m \Gamma_{\rm o} $, respectively.
By solving the coupled differential equations for radiative diffusion and wind momentum, we obtain general solutions that effectively bridge the behaviours of these limiting models.
Two key scaling results are for the terminal wind speed to escape speed, which is found to vary as $v_\infty^2/v_{\rm esc}^2 = \Gamma_{\rm o}/(1+m \Gamma_{\rm o}) -1$, and for the final observed luminosity $L_{\rm obs}$, which for all allowed steady-solutions with $m <  1 - 1/\Gamma_{\rm o}$, exceeds the Eddington luminosity, $L_{\rm obs} > L_{\rm Edd}$.
Our super-Eddington wind solutions have potential applicability for modeling phases of eruptive mass loss from massive stars,
classical novae, and the remnants of stellar mergers.
\end{abstract}


\section{Introduction and Background}
\label{sec:introduction}

The high luminosity of massive stars means that radiative forces can drive strong mass loss.
For example, the interaction of the large continuum luminosity $L_\ast$ with the bound-bound opacity of heavy minor ions is understood to lead to strong, line-driven stellar winds. 
The associated mass loss rates range up to about $10^{-5} M_\odot$\,yr$^{-1}$, 
strong enough to substantially reduce the star's mass over its evolutionary lifetime
\citep{Vink00}.
The terminal flow speeds are typically of order $v_\infty \approx 1000 - 3000$\,km\,s$^{-1}$, with a scaling that is generally a few times the escape speed $v_{\rm esc} = \sqrt{2GM/R}$ for wind initiation from a surface radius $R$ of a star with mass $M$
\citep{Mueller08}.
In Wolf-Rayet stars, the winds become modestly optically thick, with multi-line scattering leading to a wind momentum ${\dot M} v_\infty$ that can exceed the single-scattering limit $L_\ast/c$ by up to factors of ten or so
\citep{Nugis00}.
Nonetheless, even in such Wolf-Rayet stars, the total wind energy loss rate ${\dot E}_w = {\dot M} (v_\infty^2/2 + GM/R)$ is generally less than about 10\% of the stellar luminosity $L_\ast$.
In effect, because of the self-saturation of line-acceleration at large mass loss rates, such line-driven winds are {\em energetically inefficient}, with most of the core stellar energy escaping as stellar luminosity, instead of as wind energy.

The present paper explores the nature of {\em super-Eddington} winds driven by the {\em continuum} opacity
associated with free electrons, showing that these can in principle tap a much larger fraction of the total energy supplied from below.
As recently summarized by \citet{Quataert16},
there is strong evidence that massive stars undergo periods of super-Eddington energy generation/deposition, although the specific physical causes are not well understood. 
The giant eruptions of luminous blue variables (LBVs) such as Eta Carinae radiate a photon luminosity significantly exceeding the Eddington luminosity for months-decades (many dynamical times) and drive an outflow whose time-averaged kinetic power exceeds both the Eddington luminosity and probably the photon luminosity 
\citep{Smith03b,Davidson12}.
Such outbursts may dominate the total mass-loss from massive stars 
\citep[e.g.,][]{Humphreys94, Smith06, Kochanek11}.
Moreover, $\sim$10 per cent of supernova (SN) progenitors experience enhanced mass-loss in the decades to weeks prior to core
collapse (much larger than can be explained by line-driven winds).
Evidence for this powerful mass-loss includes observations of luminous outbursts that precede SNe 
\citep{Foley07,Pastorello07,Fraser13a,Fraser15,Humphreys12,Mauerhan13,Ofek13}
and mass-loss rates $\sim  0.001 - 1 \, M_\odot$\,yr$^{-1}$
inferred from observations of circumstellar interaction in Type IIn SNe 
\citep[e.g.,][]{Kiewe12,Smith14}.

As in the analysis by \citet{Quataert16} \citep[see also][]{Shen16},
the specific model explored here assumes that, within some subsurface layer,  the core stellar luminosity $L_\ast$ is supplemented by a localized, quasi-steady energy deposition rate $\Delta {\dot E}$, leading to a combined luminosity $L_{\rm o} = L_\ast + \Delta {\dot E}$ that exceeds the Eddington luminosity $L_{\rm Edd} \equiv 4 \pi GMc/\kappa_{\rm e}$.
Such a model represents a generic approach to accounting for various specific energy deposition mechanisms that could be associated with episodes of strong mass loss, including  wave deposition \citep{Piro11,Quataert12},  pre-SN core instabilities
\citep{Chen14}, and binary merger or common envelope interaction 
\citep{Ivanova13,Podsiadlowski13,Justham14}.
In some of these contexts, the energy deposition might be initially sudden, but with a magnitude that is insufficient to disrupt the overlying envelope. 
As such, following an initial prompt ejection of some fraction of the overlying envelope mass, the fallback of the overinflated, but still bound envelope can lead to a prolonged release of extra energy on a thermal timescale that is much longer than the dynamical timescale. 
If the combination of energy deposition and core luminosity exceeds  the Eddington luminosity, it implies the quasi-steady condition for  super-Eddington wind outflow explored here.

Unlike the inherent saturation of line-driving at high-densities, the continuum driving of a super-Eddington outflow can be initiated and sustained from a much deeper, denser layer, leading to much greater wind optical depths, of order $\tau \sim 10^3$ or more.
In such a very optically thick wind, the ambient radiation pressure greatly exceeds the energy density associated with the net radiative flux, $P_{\rm rad}/(F_{\rm rad}/c) \sim \tau \gg 1$.
This has led to super-Eddington models grounded in such an optically thick interior perspective
 \citep[e.g.,][]{Quataert16, Shen16},  
 which
 effectively {\em ignore} the direct driving by the radiative flux $F_{\rm rad}$.
% in favor of a radiatively dominated ($\gamma=4/3$) adiabatic wind  expansion powered by the radiative specific enthalpy $h_{\rm rad} = 4 P_{\rm rad}/\rho$.
In these models, localized heating is assumed to increase the radiative specific enthalpy $h_{\rm rad} = 4 P_{\rm rad}/\rho$; 
when this exceeds the specific gravitational binding energy $GM/R$, it leads to a simple Bernoulli solution for a radiatively dominated ($\gamma=4/3$) adiabatic wind  expansion powered by the radiative enthalpy.

In contrast to this optically thick interior approach, models grounded in a more-traditional surface wind perspective 
\citep[e.g.,][]{Owocki97, Owocki04}
assume direct {\em flux}-based continuum driving that accounts for the ``photon tiring'' reduction in radiative luminosity from the work being done to lift and accelerate the wind; but this approach {\em ignores} the dynamical effects of the large ambient radiation pressure
and the associated radiative enthalpy, precisely the terms
%precisely those 
that are important in the adiabatic models.

The aim of this paper is to reconcile and unify these two distinct previous perspectives for modeling a super-Eddington wind. 
As detailed in \S 2, our approach is grounded in solving the coupled system of differential equations for both radiative diffusion and wind momentum, 
including now both the diffusive and advective components of the radiative flux.
The non-dimensional forms of these equations (\S 2.4) are cast in terms of two key dimensionless parameters, namely the Eddington factor $\Gamma_{\rm o} \equiv L_{\rm o}/L_{\rm Edd}$, and a photon-tiring parameter $m \equiv {\dot M} GM/R L_{\rm o}$ 
\citep{Owocki97}
that characterizes that fraction of total energy input $L_{\rm o}$ needed to sustain the wind mass loss rate\footnote{
This mass loss rate is not derived explicitly here, but is effectively a free parameter, implicitly set by the density at the heating radius in terms of the gas sound speed, ${\dot M}=4 \pi R^2 \rho(R) c_{\rm sg}$.
Since this sound speed has an associated energy that is much less than the gravitational binding energy, $w_{\rm sg} = c_{\rm sg}^2/v_{\rm esc}^2 \ll 1$, it has limited dynamical effect in driving the wind. 
The  solutions derived in \S 4 thus assume the idealized limit $w_{\rm sg} \rightarrow 0$, with discussion of the effects of a small, but finite $w_{\rm sg} \approx 10^{-3} - 10^{-2}$ given in Appendix \S B.}
 ${\dot M}$ against the gravitational binding $GM/R$.
In \S 3, we show how the previous models for flux-driven vs.\ radiative-enthalpy-powered mass loss asymptotically apply in the opposite limiting regimes of, respectively, small vs.\ large values of the product $m \Gamma_{\rm o}$.
The full solutions in \S 4 show how the regimes are bridged, with the added energy initially going into radiative flux, which however  is then converted by 
radiation-pressure drag on the initial flow acceleration
into radiative-enthalpy flux that sustains the acceleration into the outer wind. 
 This section also gives key relations for how the terminal wind energy and observable luminosity scale with the parameters $m$ and $\Gamma_{\rm o}$. 
We conclude (\S 5) with a summary discussion and outlook for future work.  
The appendices give details of our solution method (\S A), provide extensions of the base analysis to account for a non-zero gas sound speed (\S B), and consider the potential role of convection in delaying the wind onset to a layer where convection becomes inefficient (\S C).

\section{General Model}
\label{sec:genmodel}

\subsection{Basic Equations}
\label{sec:basiceqns}

Following equations (4) of 
\citet{Jiang15}, we can write the 3D steady-state ($\partial/\partial t = 0$) equations for conservation of mass, momentum and energy of a gas, 
plus the diffusion equation of radiation transport, in the form, 
\beq
\nabla \cdot (\rho {\bf v}) = 0
\eeq
\beq
\nabla \cdot (\rho {\bf vv} + PI) = 
\frac{\rho \kappa}{c} ({\bf F} - {\bf v} \, E_{\rm rad}  -  {\bf v} \cdot {\tt P}_{\rm rad} ) - \rho \,  {\bf g}
\label{eq:JA2}
\eeq
\beq
\nabla \cdot [(E+P) {\bf v} +{\bf F} ] = - \rho  {\bf v} \cdot {\bf g} + {\dot q}
\label{eq:A3}
\eeq
\beq
\nabla \cdot {\tt P}_{\rm rad} =
- \frac{\rho \kappa}{c} ({\bf F}- {\bf v} \, E_{\rm rad}  -  {\bf v} \cdot {\tt P}_{\rm rad} )
\, .
\label{eq:A4}
\eeq
Here $E \equiv E_g + \rho v^2/2$  and $ E_g \equiv P/(\gamma -1)= (3/2) P$, where $P= \rho kT/\mu$ is the gas pressure in terms of mass density $\rho$, temperature $T$, and molecular weight $\mu$.

In the total energy equation (\ref{eq:A3}), we have added a volumetric heating source ${\dot q}$, with ${\bf v}$ the flow velocity, ${\bf g}$ the gravity,  and  ${\bf F}$ the observer-frame radiative flux.
For the  optically thick wind models here, we assume  the Eddington relation between radiative energy density and radiative pressure, $E_{\rm rad}=3 P_{\rm rad}$, with a correspondingly isotropic pressure tensor ${\tt P}_{\rm rad} = P_{\rm rad} \, I $. 
The radiation-pressure ``drag'' term in the momentum equation (\ref{eq:JA2}) and the radiative diffusion equation (\ref{eq:A4}) then takes the form ${\bf v} \, E_{\rm rad}  +  {\bf v} \cdot {\tt P}_{\rm rad} = 4 P_{\rm rad} \, {\bf v}$.
The opacity  $\kappa$ is taken here to be fixed to the electron scattering value, used below (see equation \ref{eq:gamodef}) to define an associated constant Eddington luminosity, $L_{\rm Edd} = 4 \pi GM c/\kappa$.

For 1D spherical symmetry with variations only in radius $r$, the above then reduce to:
\beq
{\dot M} = 4 \pi \rho v r^2 
\label{eq:Mdotdef}
\eeq
\beq
v \frac{dv}{dr} =  \frac{\kappa}{c} \left ( \frac{L}{4 \pi r^2 } - 4 v P_{\rm rad} \right ) - \frac{GM}{r^2}  - \frac{1}{\rho}  \frac{dP}{dr}
\label{eq:vdvdr}
\eeq
\beq
\frac{d}{dr}  [ {\dot M} (v^2/2+ h_g - GM/r) + L ] = 4 \pi r^2 {\dot q}
\label{eq:toten}
\eeq
\beq
\frac{dP_{\rm rad}}{dr} = - \frac{\rho \kappa}{c} \left ( \frac{L}{4 \pi r^2 } - 4 v P_{\rm rad} \right ) 
\, ,
\label{eq:raddiffr}
\eeq
where 
 ${\dot M}$ is the mass loss rate, 
 and the gas specific enthalpy $h_g \equiv (5/2) kT/\mu = (5/2) P/\rho = (5/2) c_{\rm sg}^2$, with $c_{\rm sg}$  the isothermal gas sound speed.
 
 In both the momentum equation (\ref{eq:vdvdr}) and diffusion equation (\ref{eq:raddiffr}), 
 $F= L /4 \pi r^2 $ is the  radiative flux in the stellar (observer) rest frame, while the full term in parentheses represents the diffusive flux in the flow's co-moving frame, 
 which accounts for the reduction from the advective flux $4 v P_{\rm rad} $.
 We can thus write the associated luminosities as
 \beq
 L =  L_{\rm diff} + L_{\rm adv} = L_{\rm diff} + 4\pi r^2 \, 4 v P_{\rm rad} =  L_{\rm diff} + {\dot M} h_{\rm rad}
 \, ,
 \label{eq:Ldefs}
 \eeq
where the last equality introduces the specific radiative enthalpy, $h_{\rm rad} \equiv 4 P_{\rm rad}/\rho$.
The various super-Eddington wind models in the literature differ primarily in how they treat the energy transport in equation (\ref{eq:Ldefs}).
\citet{Quataert16}  included only $L_{\rm adv}$, neglecting $L_{\rm diff}$, while the photon tiring analyses by \citet{Owocki97} and \citet{Owocki04} included $L_{\rm diff}$, but neglected $L_{\rm adv}$.

In the unified models below, we show that $L_{\rm adv}$ represents a {\em drag} on the initial acceleration by the diffusive luminosity $L_{\rm diff}$, but that the associated conversion of $L_{\rm diff}$ to advection of radiative enthalpy $L_{\rm adv}$ then powers and sustains the outer wind acceleration (see, e.g., figure \ref{fig:ldiffladvltot}).
This dual role of the advective luminosity $L_{\rm adv}$ as both an initial {\em drag} then eventual {\em driver} of the flow acceleration is one of the key, novel insights from our unified models.

\subsection{Neglect of gas enthalpy,  pressure, and sound speed}

Using the mass conservation equation (\ref{eq:Mdotdef}), we can recast the gas pressure term in the momentum equation (\ref{eq:vdvdr}) to give,
\beq
\left ( 1 - \frac{c_{\rm sg}^2}{v^2} \right ) v \frac{dv}{dr} =  \frac{\kappa}{c} \left ( \frac{L}{4 \pi r^2 } - 4 v P_{\rm rad} \right ) - \frac{GM}{r^2}  + \frac{2c_{\rm sg}^2}{r} - \frac{d c_{\rm sg}^2}{dr}
\, .
\label{eq:vdvdrcsg}
\eeq
For hot stars with mass-to-radius ratios comparable to the sun,  
%surface sound speed, $c_{\rm sg} \approx 20$\,km\,s$^{-1}$, is much smaller than the escape speed, $v_{\rm esc} \approx 600$\,km\,s$^{-1}$, 
the surface  escape speed, $v_{\rm esc} \approx 600$\,km\,s$^{-1}$ is much larger than the surface sound speed, $c_{\rm sg} \approx 20$\,km\,s$^{-1}$,
giving then a sonic energy ratio $w_{\rm sg} \equiv c_{\rm sg}^2/v_{\rm esc}^2 \approx 10^{-3} $.
At the  base of a wind outflow with optical depth $\tau \gg 1$, this increases as $w_{\rm sg} \sim T \sim \tau^{1/4}$, implying that for a very large optical depth $\tau \sim 10^{4}$, one can have a scaled sonic energy as high as $w_{\rm sg} \approx 10^{-2}$ compared to the gravitational binding energy $v_{\rm esc}^2/2 = GM/R$ at the sonic/heating radius $R$.

Within the context here of initiating a radiatively driven wind outflow from the subsurface layers of a hot, luminous star, one can thus quite generally neglect the gas specific enthalpy $h_{\rm g}$ in the energy equation (\ref{eq:toten}), as well as the sound-speed terms on the right-hand-side of the momentum equation (\ref{eq:vdvdrcsg}),
 as these have little dynamical importance in driving the wind.

If we wish to account for a smooth transition to a subsonic, nearly hydrostatic layer below the heating radius, we could optionally retain a finite value for the sound speed on the {\em left}-hand-side of (\ref{eq:vdvdrcsg}).
Appendix \S B presents results for models that include a small, but finite sonic energy $w_{\rm sg} = 10^{-3} - 10^{-2}$,
for a case in which the energy deposition is taken to be spread over narrow, but finite extent, of order the gravitational scale height in the pre-heating region.

But for the idealized model below with an arbitrarily narrow heating region, we simply take the zero-sound-speed limit even for this left-hand-side term, and use this to derive wind solutions that start from an initial speed $v(R) = c_{\rm sg} \rightarrow 0$.

\subsection{Energy and momentum requirements for outflow}

For this model of heating concentrated in a very narrow region centered on a radius $R$, i.e., $4 \pi r^2 {\dot q} = \Delta {\dot E} \, \delta (r-R)$, the total net heating 
\beq
\Delta {\dot E} (r) \equiv \int_R^r 4 \pi r'^2 {\dot q} \, dr' 
\label{eq:DeltaE}
\eeq
 is a constant for $r > R$.
The integral of the energy equation (\ref{eq:toten}) can then be solved for the luminosity for all $r>R$,
\beq
L(r) = \Delta {\dot E} (r) + L_\ast - {\dot M} \left ( \frac{v(r)^2}{2} - \frac{GM}{r} + \frac{GM}{R} \right ) 
\, ,
\label{eq:Ltire}
\eeq
where $L_\ast$ represents the underlying stellar luminosity below this heating radius;
for convenience below, we define $L_{\rm o} = \Delta {\dot E} + L_\ast$ as the constant, total energy rate input at  the lower-boundary  radius $r = R$ of the induced wind outflow.

In equation (\ref{eq:Ltire}) we have again neglected the kinetic energy associated with the initial sonic-point flow speed, since as noted above this is small compared to the gravitational binding energy $GM/R$.

The terms in (\ref{eq:Ltire}) proportional to the mass loss rate  ${\dot M}$ represent the photon tiring effect, i.e. the loss of radiative luminosity due to the work done to accelerate the flow and lift it out of the gravitational potential. Note that to ensure that the luminosity remains positive even in the case with vanishing terminal speed $v_\infty \equiv v(r \rightarrow \infty ) = 0$, we require that the mass loss rate must be below a maximal value given by
\beq
{\dot M}_{\rm max} \equiv \frac{L_{\rm o}}{GM/R} = 0.032 \frac{M_\odot}{\rm yr} \, \frac{L_{\rm o}}{10^6 L_\odot} \, \frac{R/R_\odot}{M/M_\odot} 
\, .
\label{eq:mdotmax}
\eeq
Defining a photon-tiring parameter,
\beq
m \equiv 
\frac{{\dot M}}{{\dot M}_{\rm max}} 
= \frac{{\dot M} GM}{R L_{\rm o}}
\, ,
\label{eq:mdef}
\eeq
we see that requiring $m \le 1$ represents a fundamental  energy condition that there is sufficient luminosity to drive the wind to full escape from the gravitational potential.

In terms of momentum, a further fundamental requirement is that the radiative acceleration exceed gravity, which requires that the base luminosity $L_{\rm o}$ exceed the Eddington luminosity $L_{\rm Edd} \equiv 4 \pi GM c/\kappa$.
Defining an Eddington parameter
\beq
\Gamma_{\rm o} \equiv  \frac{L_{\rm o}}{L_{\rm Edd}} = \frac{\kappa L_{\rm o}}{4 \pi GM c}
\, ,
\label{eq:gamodef}
\eeq
this momentum condition takes the form $\Gamma_{\rm o} > 1$.
For a core luminosity that is sub-Eddington, $L_\ast <  L_{\rm Edd}$,  the heating radius $R$ can\footnote{We are ignoring here the potential role that convection might have in carrying sufficient energy flux to keep the radiative luminosity sub-Eddington. Since convection can't carry the added energy to large radii, this can only  delay the onset of a super-Eddington wind to a layer where convection becomes inefficient. See Appendix \S \ref{sec:convection}.}
represent the transition from hydrostatic equilibrium to a super-Eddington wind outflow.

In line-driven stellar winds, the saturation of relatively strong lines gives the associated radiative acceleration an inverse-density dependence, $\Gamma_{\rm lines} \sim 1/\rho$; this limits the wind initiation to a relatively low-density surface layer, and makes the mass loss rate an eigenvalue, set by line opacity and stellar parameters, with small associated photon tiring values, $m < 0.01$.
In contrast, {\em continuum} driving has no such natural self-saturation (at least in 1D models without porosity effects;  see \citet{Owocki04}), so the mass loss rate has no eigenvalue, being limited only by the energy available; thus in the study here, $m$ and $\Gamma_{\rm o}$ are both treated as free parameters, physically associated with the location and level of the added heating.

Application of the luminosity from (\ref{eq:Ltire}) into the momentum equation (\ref{eq:vdvdr}) (with $P= \rho c_{\rm sg}^2 \rightarrow 0$), along with the radiation pressure equation (\ref{eq:raddiffr}), forms a coupled system of ordinary differential equations (ODE's) for  the variation of the flow speed $v$ and radiation pressure $P_{\rm rad}$ with radius $r$,
%\beq
%v \frac{dv}{dr} = 
% \frac{\kappa}{4 \pi r^2 c} 
% \left [ L _{\rm o} -  {\dot M} \left (
% \frac{v^2}{2} - \frac{GM}{r} + \frac{GM}{R}  
%+ h_{\rm rad} \right ) 
%\right ]
%- \frac{GM}{r^2}  
%\label{eq:vdvdr2}
%\eeq
%\beq
%\frac{dP_{\rm rad}}{dr} = - \frac{\rho \kappa}{4 \pi r^2 c} \left [ L_{\rm o} - {\dot M} \left ( \frac{v^2}{2} - \frac{GM}{r} + \frac{GM}{R}  + h_{\rm rad}  \right ) \right ]
%.
%\label{eq:raddiffr2}
%\eeq
\beqa
v \frac{dv}{dr} &=&
g_{\rm rad}
% \frac{\kappa}{4 \pi r^2 c} 
% \left [ L _{\rm o} -  {\dot M} \left (
% \frac{v^2}{2} - \frac{GM}{r} + \frac{GM}{R}  
%+ h_{\rm rad} \right ) 
%\right ]
- \frac{GM}{r^2}  
\label{eq:vdvdr2}
\\
%\eeq
%and
%\beq
\frac{dP_{\rm rad}}{dr} &=& 
-  \rho g_{rad}
%-\frac{\rho \kappa}{4 \pi r^2 c} \left [ L_{\rm o} - {\dot M} \left ( \frac{v^2}{2} - \frac{GM}{r} + \frac{GM}{R}  + h_{\rm rad}  \right ) \right ]
\, .
\label{eq:raddiffr2}
\eeqa
Here the radiative acceleration is given by
\beq
g_{\rm rad} \equiv \frac{\kappa}{4 \pi r^2 c} \left [ L_{\rm o} - {\dot M} \left ( \frac{v^2}{2} - \frac{GM}{r} + \frac{GM}{R}  + h_{\rm rad}  \right ) \right ]
\, ,
\label{eq:graddef}
\eeq
where the radiative-enthalpy term is related to the gas pressure by ${\dot M} h_{\rm rad}/4 \pi r^2 = 4 v P_{\rm rad}$.
The lower boundary radius $R$ represents the initiation of a super-Eddington wind, with mass loss rate ${\dot M} = 4 \pi R^2 \rho(R) c_{\rm sg}$ and flow speed $v(R)=c_{\rm sg}$.
In the idealization that $ c_{\rm sg} \rightarrow 0$, we thus have a lower boundary condition that the flow speed vanishes at this heating radius, $v(R) = 0$.

For the outer boundary at $r \rightarrow \infty$, the bracket term in the diffusion equation (\ref{eq:raddiffr2}) approaches a constant, while $\rho \sim 1/r^2$; 
this implies $dP_{\rm rad}/dr \sim 1/r^4$, and thus that the radiation pressure must vanish as $P_{\rm rad} \sim 1/r^3 \rightarrow 0$.

Recalling that $\dot{M} h_{\rm rad} \sim  v P_{\rm rad} r^2$, we thus see that in this zero-sound-speed idealization the coupling via the radiative enthalpy drag term vanishes at both the inner and outer boundaries.

\subsection{Gravitationally scaled dimensionless equations}
\label{sec:gravscaledeqns}

To solve this coupled system, it is convenient to recast it in a dimensionless form that scales the variables in terms of the gravitational escape speed $v_{\rm esc}$ and the associated gravitational escape energy, $v_{\rm esc}^2/2 = GM/R$,
\beq
w \equiv 
\frac{v^2}{v^2_{\rm esc}}  
~ ; ~ 
p \equiv \frac{4 \pi R^2 v_{\rm esc} P_{\rm rad}}{L_{\rm o}} 
~ ; ~
\eta \equiv \frac{h_{\rm rad}}{GM/R} 
\, .
\label{eqn:dlessdefs}
\eeq
Recasting also the radial independent variable as $x \equiv 1-R/r$ (which is proportional to the gravitational potential measured from the radius $R$),
the associated dimensionless forms for the coupled equations for momentum (\ref{eq:vdvdr2}) and for radiative diffusion (\ref{eq:raddiffr2}) can be written as
\beqa
\frac{dw}{dx}\!&=&\!\Gamma_{\rm o}  [ 1 - m (w +x + \eta) ]  -  1
\label{eq:dwdxeta}
\\
&=&\!\Gamma_{\rm o} \left [ 1-m(w+x)\! -\! \frac{4 p\sqrt{w}}{(1-x)^2} \right ] \! - \! 1  
\, 
\label{eq:dwdx}
\eeqa
\beqa
\frac{dp}{dx} 
&=& -  \frac{(1-x)^2}{\sqrt{w}} m \Gamma_{\rm o} \left [ 1 - m(w + x+ \eta ) \right ] 
\, 
\label{eq:dpdx0}
\\&=& -  \frac{(1-x)^2}{\sqrt{w}} m \Gamma_{\rm o} \left [ 1 - m(w + x) \right ]  + 4 m \Gamma_{\rm o} p  
\, .
\label{eq:dpdx}
\eeqa
The latter forms for each equation, viz. (\ref{eq:dwdx}) and (\ref{eq:dpdx}), thus represent the coupled system to be solved for $w(x)$ and $p(x)$ over $0 < x < 1$, with boundary conditions $w(0)=w_{\rm sg} \rightarrow 0$ and $p(1)= \eta(1) = 0$.

\section{Limiting cases}
\label{sec:limits}

Before seeking general solutions, let us consider limiting cases that recover the direct flux-driving vs.\ radiative enthalpy approaches.
As noted in the introduction, the dimensionless parameter $m \Gamma_{\rm o}$ defines two limiting regimes.
For $m \Gamma_{\rm o} \gg 1$, the drag associated with advection of radiation enthalpy is important, while for $m \Gamma_{\rm o} \ll 1$ it is not.
The next section (\S \ref{sec:tiringlim}) discusses the latter limit, while the following section (\S \ref{sec:enthalpylim}) reviews the former.  

\subsection{Direct-driving neglecting radiation-pressure drag}
\label{sec:tiringlim}

%\subsubsection{Analytic solution}

\begin{figure}
\begin{center}
\includegraphics[scale=1.03]{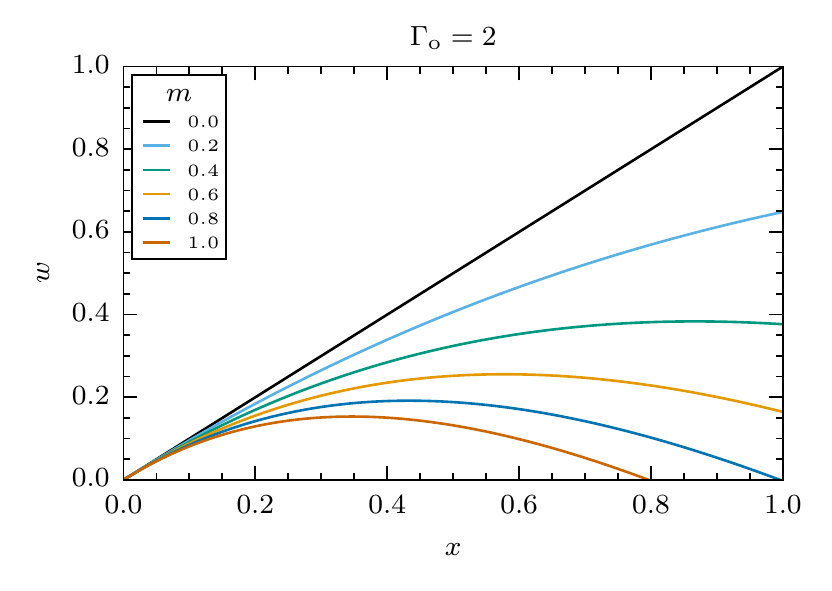}
\caption{For the case $\Gamma_{\rm o} = 2$ in the direct flux-driven model without radiative drag,  gravitationally scaled wind kinetic energy $w(x) = v^2(r)/v_{\rm esc}^2$ plotted vs.\ $x = 1-R/r $, as given by equation (\ref{eq:tiresoln}).
}
\label{fig:wxtire}
\end{center}
\end{figure}

If we simply drop the radiative-enthalpy term 
(containing $p$ or $\eta$) in the momentum equations  (\ref{eq:dwdxeta}) or (\ref{eq:dwdx}), 
then using integrating factors, we can obtain a fully analytic solution of the equation of motion (\ref{eq:dwdx}) 
\citep{Owocki97, Owocki04},
\beq
\boxed{
m (w + x) = 1 - e^{- m \Gamma_{\rm o}  x}
}
\, .
\label{eq:tiresoln}
\eeq
Figure \ref{fig:wxtire} plots $w(x)$ vs.\ $x$, for the labeled values of photon-tiring parameter $m$, in the case with $\Gamma_{\rm o} = 2$.
Note for cases with $m \gtwig 0.8$, the flow stagnates at a finite radius, i.e. at $x<1$.

For general $\Gamma_{\rm o}$, the maximum photon-tiring number $m_{\rm max}$, for which $w=0$ at $x=1$, can be computed from
\beq
m_{\rm max} = 1 - e^{- m_{\rm max} \Gamma_{\rm o}  } ~~ \rightarrow ~~ m_{\rm max} = 1+ \frac{ W \left (-\Gamma_{\rm o} \, e^{-\Gamma_{\rm o}} \right ) }{\Gamma_{\rm o}} 
\, ,
\label{eq:mmax}
\eeq
where $W$ is the product-log (or Lambert) function.
For $\Gamma_{\rm o} \gtrsim 1$, $m_{\rm max} \lesssim 1$.

For flows with $m \le m_{\rm max}$, the ratio of observed luminosity $L_{\rm obs}$ over base luminosity $L_{\rm o}$ scales as
\beq
\frac{L_{\rm obs}}{L_{\rm o}} 
= 1 - m(w(1)+1) = e^{-m \Gamma_{\rm o}} 
\, .
\label{eq:Lobstire}
\eeq
For the maximal photon-tiring case, $m=m_{\rm max}$, this gives 
\beq
\frac{L_{\rm obs}}{L_{\rm o}} = e^{- \Gamma_{\rm o}  - W \left (-\Gamma_{\rm o} \, e^{-\Gamma_{\rm o}} \right )} \rightarrow 0 ~~ {\rm as} ~ \Gamma_{\rm o} \rightarrow \infty
\, .
\label{eq:Lobstiremax}
\eeq

%\subsubsection{Self-consistency of neglecting radiative drag}

Let us next examine the self-consistency of neglecting the radiative drag, i.e., the advection of radiative enthalpy.
For this, note that the ratio of the advection of radiative enthalpy to the diffusive radiative flux has the scaling
\beq
\frac{4 v P_{\rm rad}}{c |dP_{\rm rad}/d\tau |} 
\approx  \frac{4 v P_{\rm rad}}{c P_{\rm rad}/\tau} 
= 4  \tau v/c
\, .
\label{eq:pradvsfrad}
\eeq
We thus need to compute
\beqa
4 \tau(r) \frac{v(r)}{c} 
&=& \frac{\kappa {\dot M} v(r) }{\pi  c} \int_r^\infty \frac{dr'}{v(r') r'^2} 
\nonumber
\\
&=& 4 m \Gamma_{\rm o}  \sqrt{w(x)} \int_x^1 \frac{dx'}{\sqrt{w(x')} }
\, ,
\label{eq:tauvbc}
\eeqa
where $m \Gamma_{\rm o} = \kappa {\dot M}/4 \pi R c$ provides the overall scale for this optical depth.

In the weak-photon-tiring limit $m \ll 1$,  the velocity has the simple solution $\sqrt{w(x)} =  \sqrt{( \Gamma_{\rm o} - 1 ) x} $, giving
\beq
4 \tau v/c= 8 m \Gamma_{\rm o}  \sqrt{x} \left ( 1-\sqrt{x} \right ) ~~ ; ~~ m \ll 1
\, ,
\label{eq:mll1}
\eeq
which becomes zero at both the surface, $x=0$, and at large radii, $x=1$, with a peak value of $2 m \Gamma_{\rm o}$ at $x=1/4$.

\begin{figure}
\begin{center}
\includegraphics[scale=1.03]{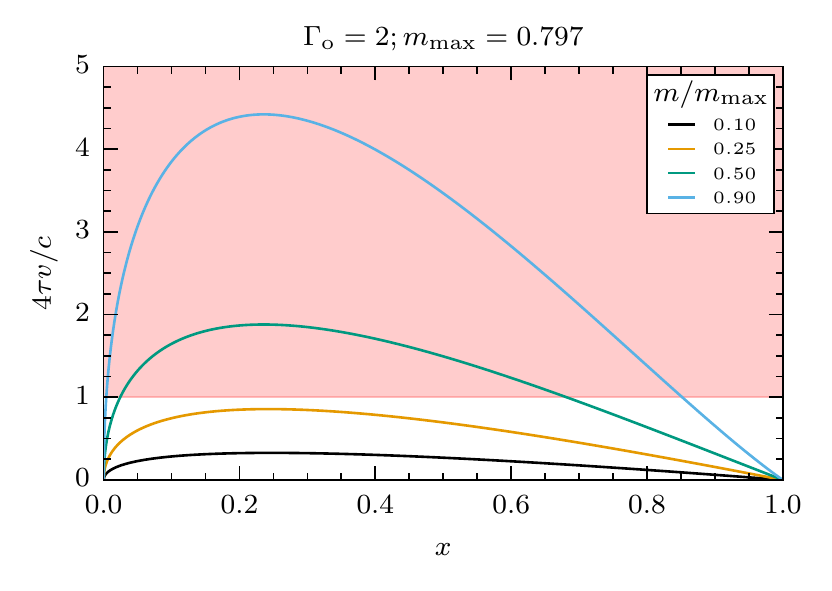}
\caption{For the case $\Gamma_{\rm o} = 2$, the ratio of 
the advective to diffusive flux in solutions (equation \ref{eq:tiresoln}) that neglect the former,
plotted vs.\ $x=1-R/r$ for various ratios of photon-tiring parameter $m$ to its maximum value $m_{\rm max} = 0.797$. In the entire red region, $4 \tau v/c > 1$, implying that 
radiation advection should be dynamically important, so should not be ignored.}
\label{fig:taubeta4}
\end{center}
\end{figure}

Figure \ref{fig:taubeta4} plots $4\tau v/c$ vs. $x$ for the case $\Gamma_{\rm o} = 2$, and for various ratios of the photon-tiring parameter to 
the maximum value given by equation (\ref{eq:mmax}).
In the entire red region, $4 \tau v/c > 1$, implying that radiative drag should be dynamically important, and so should not be ignored.
This issue becomes even more problematic for models with larger $\Gamma_{\rm o}$, and ratios $m/m_{\rm max}$ close to unity.

But at the base (and far from the star), the drag becomes small, implying one can still initiate the wind outflow with this formalism.
We return to this point in our discussion of full solutions in \S  \ref{sec:fullsoln}.

\subsection{Radiative enthalpy neglecting diffusive radiative flux}
\label{sec:enthalpylim}

%\subsubsection{Analytic solution based on Bernoulli equation}

In the opposite limit $m \Gamma_{\rm o} \gg 1$, we find that  the radiative diffusion equation (\ref{eq:dpdx}) takes the form
\beq
- \frac{(1-x)^2}{\sqrt{w}} \left [ 1 - m(w + x) \right ]  + 4 p  = \frac{1}{m \Gamma_{\rm o}} \frac{dp}{dx} \rightarrow 0 
%~ ; ~ m \Gamma_{\rm o} \gg 1
\, .
\label{eq:plima}
\eeq
Setting the pressure gradient term on the right-hand-side to zero, we can recast equation (\ref{eq:plima}) in the form of a Bernoulli equation for wind energy $w$ in terms the spatial coordinate $x$,
\beq
w + x + \eta =
w + x + \frac{4p}{m}  \frac{\sqrt{w}}{(1-x)^2}   
= \frac{1}{m}
\, .
\label{eq:plimb}
\eeq
Using the definitions in (\ref{eqn:dlessdefs}), conversion back to dimensional form gives 
\beq
\frac{v^2}{2}  - \frac{GM}{r} + h_{\rm rad} =   
\frac{L_{\rm o}}{{\dot M}} - \frac{GM}{R} 
\, .
\label{eq:Bes2}
\eeq
This conservation of energy (Bernoulli) equation was central to the super-Eddington wind model by \citet{Quataert16} (see their equations 8 and 9).
The enthalpy in (\ref{eq:Bes2}) can be written in terms of a {\em radiative} sound speed, $h_{\rm rad} = 4 P_{\rm rad}/\rho = 4 c_{\rm sr}^2$.
At the associated radiative sonic radius $r_{\rm sr}$, we require $v_{\rm sr}^2 = (4/3) c_{\rm sr}^2$ and $c_{\rm sr}^2 = (3/8) GM/r_{\rm sr}$
\citep[][see their equation 12]{Quataert16},
which when applied to equation (\ref{eq:Bes2}) gives
 \beq
\frac{L_{\rm o}}{{\dot M}} - \frac{GM}{R} 
= \frac{3}{4} \frac{GM}{r_{\rm sr}} 
= \frac{3}{2} v_{\rm sr}^2
\, .
\label{eq:Bes3}
\eeq
Dividing through by $GM/R$, this can be recast in dimensionless terms,
\beq
\frac{1-m}{m} =  \frac{3}{4} \, (1-x_{\rm sr} ) = 3 w_{\rm sr}
\, ,
\label{eq:Bes4}
\eeq
which can be readily solved to give
\beq
 w_{\rm sr} 
  = \frac{1-m}{3m} ~~ ; ~~  x_{\rm sr} 
 = \frac{7m-4}{3m}
 \, .
 \label{eq:wsxs}
 \eeq
Using the fact that $\eta \sim  \rho^{1/3}  \sim (1-x)^{2/3}/w^{1/6}$,
the dimensionless Bernoulli equation (\ref{eq:plimb}) can then be recast as
\beq
\boxed{
w + x + 
2 \left ( \frac{1-m}{m}  \right )
 \left ( \frac{1-x}{1-x_{\rm sr}} \right )^{2/3}  \left ( \frac{w_{\rm sr}}{w} \right )^{1/6}
= \frac{1}{m}
}
\, .
\label{eq:plimc}
\eeq
Note that once the requirement $m \Gamma_{\rm o} \gg 1$ for enthalpy-driven flow is satisfied, such enthalpy solutions {\em depend only on $m$}, and are {\em independent} of the particular value of $\Gamma_{\rm o}$.

 Setting $x=1$ in equation (\ref{eq:plimc}) shows that the terminal flow energy for this enthalpy model is given simply by 
\beq
w(1)=1/m -1 
\, .
\eeq
For all $m$,  evaluations of the Bernoulli solution (\ref{eq:plimc}) give a  terminal wind energy that is 3 times the wind energy at the sonic point, $w(1)=3 w_{\rm sr}$.

For the fully tired case $m \rightarrow 1$, the sonic-point speed vanishes, $ w_{\rm sr} \rightarrow 0$, while $x_{\rm sr} \rightarrow 1$, implying that the sonic point recedes to large radii, $r_{\rm sr} \rightarrow \infty$.
 For a more moderate case in which the sonic radius is at twice the heating radius, $r_{\rm sr}/R = 2$ (and so $x_{\rm sr}=1/2$), we require $m=8/11= 0.73$. In this case, $w_{\rm sr} = 1/8 = 0.125 $. with a base flow energy $w(0)=0.073$.
 
Somewhat more surprising is that for $m  < 4/7$, $x_{\rm sr}$ becomes {\em negative}, indicating
that for these cases, the {\em initial flow at the heating radius must already be supersonic} i.e., $w(0) > w_{\rm sr}$, in the sense of this radiative sound speed, $c_{\rm sr}$, which is much higher than the gas sound speed $c_{\rm sg}$ discussed \S \ref{sec:basiceqns}.
This stems in part from the simplification that the heating all takes place in a vanishingly small region at $r = R$.   
But it also reflects the fact that, as a purely algebraic solution, this Bernoulli approach does not have a boundary condition that can insure low outflow speed from the heating radius $R$.
As discussed below (see, e.g., figure \ref{fig:wvsx-g100-m468} and associated text), this apparent inconsistency of the strict radiative enthalpy model is rectified by including the diffusive radiative flux. 
In models with large but finite $m \Gamma_{\rm o}$, this leads to an initial boundary layer just above the heating radius, where the flow is accelerated from small values and the diffusive radiative flux is converted into advection of radiative enthalpy.

%\subsubsection{Enthalpy scaling for observed luminosity}
\label{sec:Lobsenth}

Within this enthalpy model, the observed luminosity can be estimated from the common value of the advective vs.\ diffusive luminosity at an outer diffusion radius, which by equation 
(\ref{eq:pradvsfrad}) occurs near where $4 \tau v/c \approx 1$.
Using equation (\ref{eq:mll1}), and assuming that the flow speed at this radius is near its terminal value $w_1 = 3 w_{\rm sr} $,
we find that this diffusion location is set by $x_{\rm d} \approx 1 - 1/(4 m \Gamma_o )$.
Applying this in the scaling for the advective luminosity with enthalpy, we obtain
\beqa
\frac{L_{\rm obs}}{L_{\rm o}} &\approx&  \left [ \frac{L_{\rm adv}}{L_{\rm o}} \right]_{\rm d}
= m  \eta_{\rm d} 
\nonumber
\\
 &\approx& 2(1-m)  \left ( \frac{1-x_{\rm d}}{1-x_{\rm sr}} \right )^{2/3}  \left ( \frac{w_{\rm sr}}{w_1} \right )^{1/6}
 \nonumber
 \\
 & \approx& 
  \, \left ( \frac{1-m}{\Gamma_{\rm o}^2} \right )^{1/3}
\, ,
\label{eq:Lobsenth}
\eeqa
where 
the second equality assumes $w_{\rm d} \approx w_1$, and the last  form replaces an order-unity collection of constants with unity.
In terms of the Eddington luminosity, the scaling in equation (\ref{eq:Lobsenth}) can be written as
\beq
L_{\rm obs} \sim L_{\rm Edd}  \left( \frac{L_{\rm o}}{L_{\rm Edd}} \right )^{1/3}
\, ,
\label{eq:LobsQ41}
\eeq
%$L_{\rm obs} \sim L_{\rm Edd} (L_{\rm o}/L_{\rm Edd})^{1/3}$, 
which agrees with equation (41) of  \citet{Quataert16} \citep[see also][]{Meier82, Begelman83, Shen16};
but the full result (\ref{eq:Lobsenth}) now includes an explicit dependence on the photon-tiring parameter $m$.

\begin{figure}
\begin{center}
\includegraphics[scale=1.35]{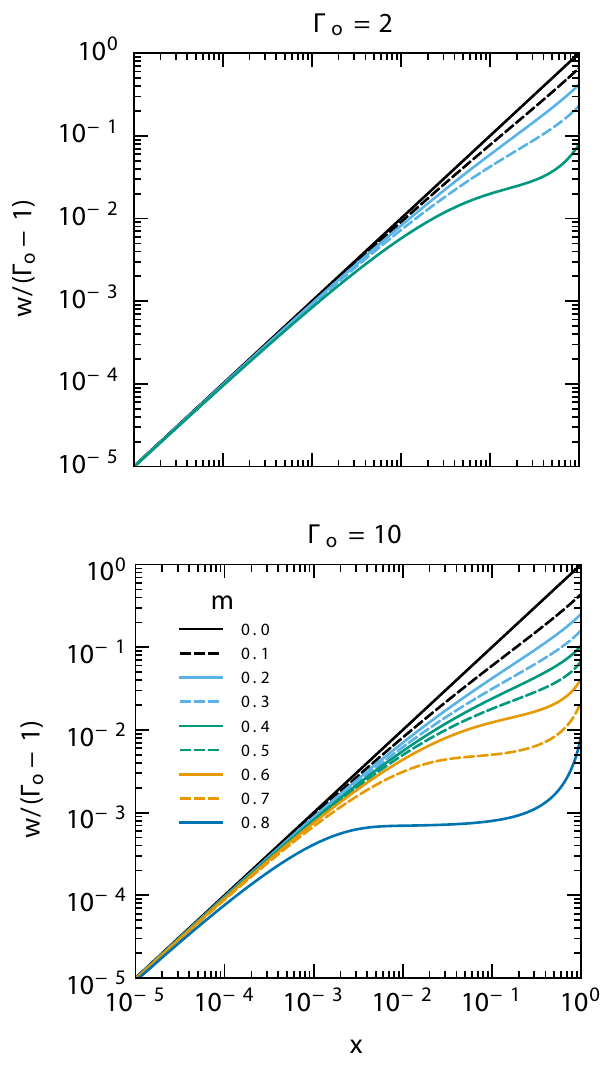}
\caption{
Full solutions for scaled wind energy $w/(\Gamma_{\rm o} -1)$ plotted vs. spatial coordinate $x$ for Eddington parameters  $\Gamma_{\rm o} =$ 2 (top) and 10 (bottom), 
each with various values of photon-tiring parameter $m$.
Compared with the direct acceleration of the no photon-tiring case $m=0$ (straight blue lines), each case with increasing photon-tiring shows a slower acceleration, and lower terminal speed.
%As the photon-tiring approaches the maximum value $m_{\rm max} = 1 - 1/\Gamma_{\rm o}$ for each case, the wind acceleration shows a stagnation region, with a plateau in wind energy, before acceleration resumes near the outer boundary.
}
\label{fig:w-vs-x-Gs}
\end{center}
\end{figure}

\section{Full solution of coupled ODE's}
\label{sec:fullsoln}

Let us next develop and examine full solutions of the coupled system of equations for wind momentum (\ref{eq:dwdx}) and radiation pressure (\ref{eq:dpdx}).

In the limit $m \Gamma_{\rm o} \ll 1$,  standard differential equation solvers give stable solutions with a form that confirm quite well the analytic direct flux-driven solution (\ref{eq:tiresoln}).
However, in the opposite limit $m \Gamma_{\rm o} \gg 1$ of an enthalpy-powered-flow,
the fact that the differential equation (\ref{eq:dpdx}) for the radiative flux reduces to the algebraic, Bernoulli form (\ref{eq:plimb}) implies that the general coupled ODE system (\ref{eq:dpdx}) and (\ref{eq:dwdx}) becomes very {\em stiff} in this limit.

As detailed in  Appendix \S \ref{sec:inoutsoln}, solutions in this general case thus require a more careful approach that matches outward integrations from the lower boundary with inward integrations from the outer boundary.
The remainder of this section presents results of these full solutions of the coupled system.

\subsection{Spatial variation of flow quantities}

For Eddington parameters $\Gamma_{\rm o} =$ 
%2, 10, and 100, the left, middle, and right panels
2 and 10, the upper and lower panels
of figure \ref{fig:w-vs-x-Gs} plot the spatial variation of wind energy scaled by the Eddington parameter, $w(x)/(\Gamma_{\rm o} - 1)$, 
each with a range of photon-tiring parameter $m$ below their respective maximum value $m_{\rm max}=1 - 1/\Gamma_{\rm o}$.
Compared with the direct acceleration of the no photon-tiring case $m=0$ (straight blue lines), each case with increasing photon-tiring shows a slower acceleration, and lower terminal speed.

\begin{figure}
\begin{center}
\includegraphics[scale=1.03]{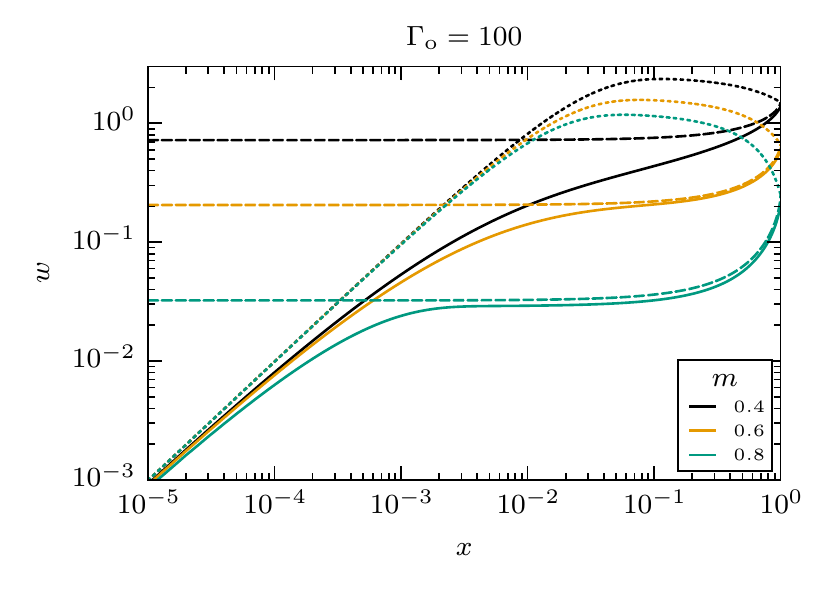}
\caption{
For the strongly super-Eddington case  $\Gamma_{\rm o}=100$, comparison of the spatial variation of wind kinetic energy $w$ (on a log-log scale) for photon-tiring parameters $m=0.4$, 0.6, and 0.8 (respectively blue, purple, red curves). 
Note how the full solutions for $w$ (solid curves) bridge the variations of the direct flux-driven model without radiation drag (dotted curves) and the enthalpy model (dashed curves).
}
\label{fig:wvsx-g100-m468}
\end{center}
\end{figure}

For the strongly super-Eddington case  $\Gamma_{\rm o}=100$, figure \ref{fig:wvsx-g100-m468}  compares the spatial variation of wind kinetic energy $w$ (on a log-log scale) for photon-tiring parameters $m=0.4$, 0.6, and 0.8 (respectively blue, purple, red curves). 
This shows how the full solutions for $w$ (solid curves) effectively ``bridge'' the variations of the direct flux-driven model without radiation drag (dotted curves) and the enthalpy model (dashed curves).

\begin{figure}
\begin{center}   
\includegraphics[scale=1.03]{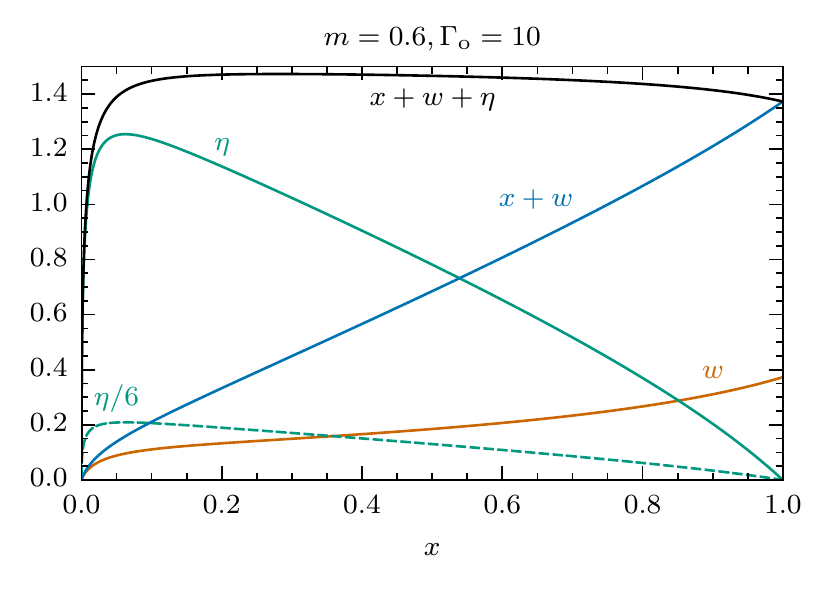}
\caption{Comparison of spatial variation of various flow variables for a sample case with $\Gamma_{\rm o}=10$ and $m=0.6$. 
The initial near-base increase of the  flow energy $w$ (red curve) comes from the direct flux-driving by the super-Eddington radiation with $\Gamma_{\rm o} = 10$, but the radiation drag also leads to a concomitant near-base increase in the specific enthalpy $\eta$ (green curve). 
The total specific energy $x+w+\eta$ (black curve)  remains nearly flat after this initial buildup, with the rise in wind kinetic + potential energy $x+w$ (blue curve) powered by the drop in radiative enthalpy $\eta$.
The dashed green curve for $\eta/6$ represents the square of the adiabatic radiative sound speed.}
\label{fig:flow-vars-lin}
\end{center}
\end{figure}

Figure \ref{fig:flow-vars-lin} compares the spatial variation of various flow variables for a sample case with $\Gamma_{\rm o}=10$ and $m=0.6$.
The initial near-base increase of the flow energy $w$ (red curve) comes from the direct flux-driving by the super-Eddington radiation with $\Gamma_{\rm o} = 10$;
but the radiation drag also leads to a concomitant near-base increase in the specific enthalpy $\eta$ (purple curve).
The total specific energy $w+\eta+x$ (black curve) remains nearly flat following the initial buildup, with the steady growth in total wind energy $w+x$ (blue curve) effectively powered by the marked drop in radiative enthalpy (purple curve).
The overall result is a slower, more extended acceleration (red curve) than occurs in weak photon-tiring models with small $m$.

The dashed purple curve for $\eta/6$ represents the square of the adiabatic radiative sound speed.
 The intersection of this with the wind energy $w$ (red curve) represents the radiative sonic point, given here by $x_{\rm sr} \approx 0.36$. 
For $m=0.6$ but larger $\Gamma_{\rm o}$, this sonic point approaches the analytic value $x_{\rm sr} = (7m-4)/(3m) = 0.111$ predicted by equation (\ref{eq:wsxs}) for the radiative enthalpy limit ($m \Gamma_{\rm o} \gg 1$).

\subsection{Spatial variation of luminosity components}

\begin{figure}
\begin{center}
\includegraphics[scale=1.03]{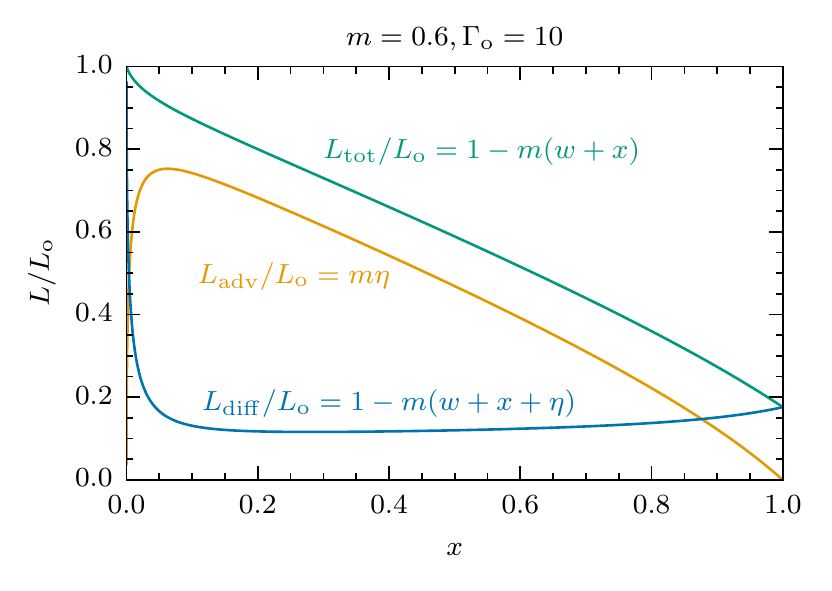}
\caption{
Spatial variation of diffusive luminosity (blue curve),  
advective luminosity  (orange curve), and
total luminosity  (green curve) 
for standard model with $m =0.6$ and $\Gamma_{\rm o} = 10$.
}
\label{fig:ldiffladvltot}
\end{center}
\end{figure}

For this same  sample case with $\Gamma_{\rm o}=10$ and $m=0.6$, figure \ref{fig:ldiffladvltot} compares the spatial variation of various components of luminosity.
The diffusive component $L_{\rm diff}$ drops sharply from the base, due primarily to losses to the radiative enthalpy $\eta$, which leads to  a sharp initial rise in the advective luminosity $L_{\rm adv} = L_{\rm o} m \eta$, followed by a decline as the enthalpy is used to help sustain the outflow.
The enthalpy thus acts as a storage conduit for the overall decline of total luminosity $L_{\rm tot}  = L_{\rm diff} + L_{\rm adv}$ due to work done in lifting and accelerating the flow.

The advective and diffusive luminosities are equal to each other at the diffusion radius $r_{\rm d }= R/(1-x_{\rm d}) \approx 4 m \Gamma_{\rm o} R$, where $\tau \sim c/v$.   
This estimate is reasonably consistent with the numerical result in Figure \ref{fig:ldiffladvltot} and becomes yet more accurate for larger values of $m \Gamma_{\rm o}$.

\subsection{Wind optical depth}

The radial optical depth of the wind is computed from
\beq
\tau (x) = \int_{r(x)}^\infty \kappa \rho(r') \, dr' 
= \tau_\ast
\int_x^1 \frac{dx'}{\sqrt{w(x')}}
\, ,
\label{eq:taux}
\eeq
where for the lower bound of the first integral we recall that $r(x)=R/(1-x)$;
the second equality introduces an optical depth scaling factor,
\beq
\tau_\ast \equiv \frac{\kappa {\dot M}}{4 \pi R \, v_{\rm esc}} =  m \Gamma_{\rm o} \frac{c}{v_{\rm esc}}
\, .
\label{eq:tauast}
\eeq
For a mass-to-radius ratio $M/R$ of the order the solar value, we have $c/v_{\rm esc} \approx 500$, giving for this case $\tau_\ast \approx 500 m \Gamma_{\rm o} = 3000$. From numerical integration, we find the optical depth ratio at the wind base is typically order unity,
 e.g.\ for the standard case $m=0.6; ~\Gamma_{\rm o}=10$ we find $\tau (0)/\tau_\ast \approx 2.4 $, 
implying then a very large base optical depth, $\tau (0) \approx 7000$.

\begin{figure}
\begin{center}
\includegraphics[scale=1.03]{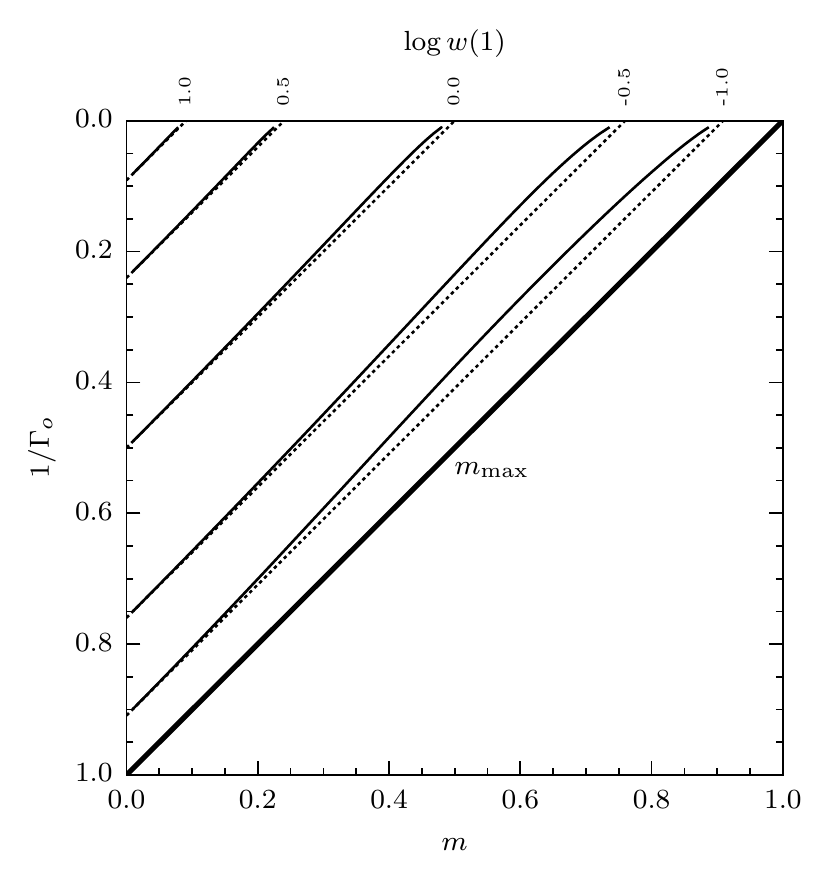}
\caption{
Logarithm of the terminal wind energy, $\log w(1)$, from full solutions of the coupled ODE's for energy $w$ and pressure $p$, plotted as contours vs.\ $1/\Gamma_{\rm o}$ and $m$.
The dotted curves compare the linear ``bridging'' form 
%(see figure \ref{fig:wacontvsmgi}) 
for the terminal wind energy $w (1) \approx  \Gamma_{\rm o}/(1+m \Gamma_{\rm o} )-1$ between the 
limiting cases $w (1) = \Gamma_{\rm o}-1$ for $m \Gamma_{\rm o} \ll 1$ and $w(1) \approx -1+1/m$ for $m \Gamma_{\rm o} \gg 1$.
The black diagonal line shows the maximum tiring parameter $m_{\rm max} \equiv 1-1/\Gamma_{\rm o}$;
%above which (gray shaded region) 
in the region to right and below this line,
there are no solutions reaching to large distance, $x \rightarrow1$.
}
\label{fig:term_contour-stanedit}
\end{center}
\end{figure}

Even for the weaker super-Eddington case $\Gamma_{\rm o} = 2$ with weaker photon-tiring  $m=0.1$, we find $\tau (0)/\tau_\ast \approx 2.3$,  still implying a quite large base optical depth $\tau (0) = 2.3 \, m \Gamma_{\rm o} \, 500 = 230$.

For small photon-tiring parameter $m \ll 1/\Gamma_{\rm o}$, equation (\ref{eq:tiresoln}) gives $w(x) \approx (\Gamma_{\rm o} -1) x$, which when applied in equation (\ref{eq:taux}) gives a base optical depth $\tau(0) = 2 \tau_\ast/\sqrt{\Gamma_{\rm o} -1}$. 
Using (\ref{eq:tauast}), this can be solved for the tiring parameter that would have unit optical depth, 
\beq
m_{\tau(0)=1} = \frac{\sqrt{\Gamma_{\rm o} -1}}{2 \Gamma_{\rm o}} \, \frac{v_{\rm esc}}{c}
\approx \frac{0.001}{\sqrt{\Gamma_{\rm o}}}
\, ,
\label{eq:mtau1}
\eeq
where the latter approximation applies for the solar $M/R$ and $\Gamma_{\rm o}$ more than order unity.

The upshot is that, 
apart from such very low photon-tiring parameter values, these super-Eddington wind solutions are indeed generally quite optically thick.

\subsection{Scaling laws for terminal  wind speed and observable luminosity}

Let us next examine the scaling of the terminal wind kinetic energy $w(1) = v_\infty^2/v_{\rm esc}^2$.
Figure \ref{fig:term_contour-stanedit} shows contours of the full solution for $\log w(1)$ as a function of $m$ and $1/\Gamma_{\rm o}$.
The dotted lines compare 
%the linear interpolation scaling introduced in figure \ref{fig:wacontvsmgi}, which 
a linear interpolation scaling that
bridges between the limiting cases 
$w(1) = \Gamma_{\rm o}-1$ for $m \Gamma_{\rm o} \ll 1$ and $w(1) \approx -1+1/m$ for $m \Gamma_{\rm o} \gg 1$, 
viz.\
\beq
w (1) \approx
\boxed{
\frac{ \Gamma_{\rm o}}{1+m \Gamma_{\rm o} } -1
\approx \frac{v_\infty^2}{v_{\rm esc}^2}
}
\, .
\label{eq:w1b}
\eeq

This scaling for the observable luminosity applies for moderate Eddington parameters $\Gamma_{\rm o} < 10$.
For any given $\Gamma_{\rm o}$, there is a maximum photon-tiring parameter $m_{\rm max} = 1 - 1/\Gamma_{\rm o}$, for which $w(1)=0$.
For $m \rightarrow m_{\rm max}$ one finds $\Gamma (1) =1$, implying that the observed luminosity would just be equal to the Eddington luminosity $L_{\rm obs} = L_{\rm Edd}$.
%The upshot is that 
Thus in this full model that accounts for radiation-pressure drag, the observed luminosity is always equal to or greater than the Eddington value.

Equation (\ref{eq:w1b}) also suggests a simple scaling for the ratio of the terminal (observable) luminosity to the lower boundary input value,
\beq
 \frac{\Gamma (1) }{\Gamma_{\rm o}} = 1 - m(w(1) + 1) \approx 
\boxed{
\frac{1}{1+m \Gamma_{\rm o}}  
\approx \frac{L_{\rm obs}}{L_{\rm o}}
}
\, .
\label{eq:LobsLo}
\eeq

For larger $\Gamma_{\rm o} > 10$ and $m \lesssim m_{\rm max} \approx 1-1/\Gamma_{\rm o}$, the observed luminosity follows the enthalpy scaling (\ref{eq:Lobsenth}) derived in \S \ref{sec:Lobsenth}.
As noted by  \citet[][see their equation 41]{Quataert16}, in the enthalpy limit  the observed luminosity can exceed the Eddington value by an even greater factor, $L_{\rm obs}/L_{\rm Edd} \sim  \, \Gamma_{\rm o}^{1/3}$  \citep[see also][]{Meier82, Begelman83, Shen16}.

\begin{figure}
\begin{center}
\includegraphics[scale=1.03]{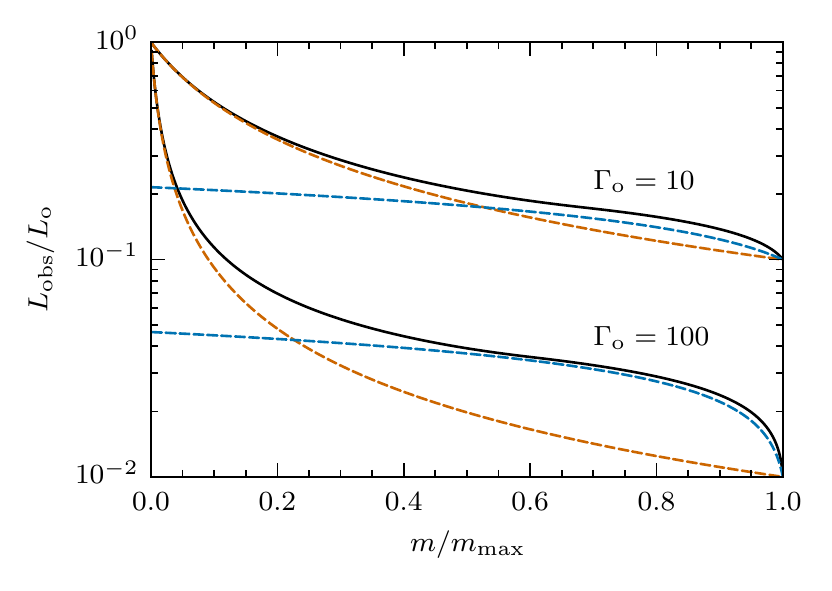}
\caption{
For Eddington parameters $\Gamma_{\rm o} =$ 10  (top curves) and 100 (bottom curves), the observed luminosity ratio $L_{\rm obs}/L_{\rm o}$ plotted vs.\ $m/m_{\rm max}$, for the full solution (black curves),
and the two scaling relations  (\ref{eq:Lobsenth}) (dotted blue curves)  and (\ref{eq:LobsLo}) (red dashed curves).
}
\label{fig:Lobsg10g100}
\end{center}
\end{figure}

For the cases $\Gamma_{\rm o} =$ 10 and 100, figure \ref{fig:Lobsg10g100} compares $L_{\rm obs}/L_{\rm o}$ vs. $m/m_{\rm max}$ from the full solution (solid black curves) with scaling results (\ref{eq:Lobsenth}) (dotted blue curve)  and (\ref{eq:LobsLo}) (red dashed curve).
For both $\Gamma_{\rm o}$ values, the enthalpy scaling (\ref{eq:Lobsenth})  provides a good fit for large $m \lesssim m_{max}$, but fails to produce the upturn toward $L_{\rm obs} \rightarrow L_{\rm o}$ in the weak photon-tiring limit $m \rightarrow 0$.
For the case with $\Gamma_{\rm o} =10$,
the simple scaling (\ref{eq:LobsLo})  provides a reasonably good fit for all $m$;
but for $\Gamma_{\rm o} = 100$, this scaling underestimates $L_{\rm obs}/L_{\rm o}$ for large $m \lesssim m_{\rm max}$.
For both $\Gamma_{\rm o}$, the two scalings intersect at an intermediate tiring number $m_{\rm b} \approx  (0.2 - 0.5) \, m_{\rm max} $.
Thus a rough law connecting the two limits would be to use the simple scaling (\ref{eq:LobsLo}) for $m \le m_{\rm b}$, and
the enthalpy form (\ref{eq:Lobsenth}) for $m > m_{\rm b}$.
Comparing equations (\ref{eq:Lobsenth}) and (\ref{eq:LobsLo}), we find that $m_{b} \approx \Gamma_{\rm o}^{-1/3}$, in good agreement with figure \ref{fig:Lobsg10g100}.

\section{Summary and Future Outlook}
\label{sec:summary}

This paper derives semi-analytic solutions for a super-Eddington wind for an idealized model with a steady-state deposition of energy centered narrowly on a near-surface radius $R$.
In addition to the momentum requirement that the total base luminosity $L_{\rm o}$ exceed the Eddington limit, $\Gamma_{\rm o} \equiv L_{\rm o}/L_{\rm Edd} > 1$,  there is an energy requirement that this luminosity should also be sufficient to sustain the mass loss against the gravitational binding, implying that the photon-tiring parameter $m \equiv {\dot M} GM/R L_{\rm o} < 1-1/\Gamma_{\rm o}$.

A key motivation for the analysis here is to reconcile and unify previous super-Eddington wind models that were grounded in surface \citep[e.g.,][]{Owocki04} vs.\  interior \citep[e.g.,][]{Quataert16} perspectives, and so made divergent assumptions about the importance of the radiative flux vs. radiative enthalpy in the wind driving.
The analysis in \S \ref{sec:limits} shows that these apply in opposite limits of small vs.\ large values of $m \Gamma_{\rm o}$, respectively.

Using methods outlined in Appendix \S A, solutions of the coupled system of equations for  wind momentum (\ref{eq:dwdx}) and radiative diffusion (\ref{eq:dpdx}) give general results (\S \ref{sec:fullsoln}) that show how the limiting regimes are bridged; the added energy near the base of the wind initially goes into radiative flux, which however  is then converted 
into radiative enthalpy that sustains the acceleration in the outer wind.  
Two key scaling results for these full wind solutions regard the wind terminal speed, given by equation (\ref{eq:w1b}), and  the observable luminosity, given by 
equations (\ref{eq:Lobsenth}) and (\ref{eq:LobsLo}).
In particular, the observable radiative luminosity is found always to be at least the Eddington value, $L_{\rm obs} \ge L_{\rm Edd}$, even in the strong photon-tiring limit $m \rightarrow m_{\rm max} = 1-1/\Gamma_{\rm o}$, for which $v_\infty \rightarrow 0$.

The mass loss rate is not derived explicitly here, but is effectively a free parameter, implicitly set by the density at the heating radius in terms of the gas sound speed, ${\dot M}=4 \pi R^2 \rho(R) c_{\rm sg}$.
Since this sound speed has an associated energy that is much less than the gravitational binding energy, $w_{\rm sg} = c_{\rm sg}^2/v_{\rm esc}^2 \ll 1$, it has limited dynamical effect in driving the wind. The  solutions derived in \S 4 thus assume the idealized limit $w_{\rm sg} \rightarrow 0$.
 Appendix \S B shows explicitly that including a small, but finite $w_{\rm sg} \approx 10^{-3} - 10^{-2}$ has little effect on the solutions above the gas sonic point.
The fact that ${\dot M} \sim \rho (R) $ implies that heating must be concentrated near the surface (where the density is much less than in the interior), in order for $m < 1$, i.e., in order to overcome the star's gravitational binding 
(see \citet{Quataert16} for more details).

One key issue not directly addressed by the analysis here regards the response of the envelope to heating at a deep and dense enough layer that a directly induced mass loss rate would exceed the photon-tiring limit, $m > 1$. 
For this case, 1D time-dependent hydrodynamical simulations 
\citep{vanMarle09,Quataert16}
show an initial launching of mass outflow that eventually stagnates. 
In the simulations by \citet{vanMarle09}, 
the re-energization of the radiation during fallback leads to repeated phases of upflow and downflow, with a net outward lifting of a reduced mass flux to full escape.  
These simulations did not, however, take any account of the advection of radiation and the associated radiation-pressure drag-effects that are expected 
for such an optically thick outflow. 

Perhaps more fundamentally, the 1D shells in this model would be expected to break up into clumps, e.g.\ by Rayleigh-Taylor instabilities, in a more realistic multi-dimensional model.
One potentially key effect would be to make the medium ``porous'' \citep{Shaviv98,Begelman01}, with then a reduced radiative driving that might regulate the mass loss to a rate that can be energetically sustained \citep{Owocki04}.
Another possibility, considered in a preliminary way in Appendix \S C, is that the heating induces convection, with an associated energy transport in deeper layers that is 
%sufficient to carry enough of 
efficient enough in carrying
the added energy to keep the radiative flux below the Eddington limit. The analysis in Appendix \S C indicates that the details of the convective saturation will determine whether this can effectively delay the onset of mass loss to a lower-density layer for which 
$m < m_{\rm max} =1 - 1/\Gamma_{\rm o}$, so that the initiated mass loss is energetically sustainable.

A key issue for future work will thus be to carry out multi-dimensional simulations using a radiation-hydrodynamics code that can both  track the convective energy transport, and account for any porous reduction in the opacity, associated with extensive density inhomogeneity from failed outflow and subsequent fallback.  
A promising prototype is provided by the recent 3D simulations by 
\citet{Jiang15} of radiative driving near the iron opacity bump in massive stars that are just below the electron-scattering Eddington limit.

Finally, there is of course a need to connect the idealized super-Eddington wind results here to models with a specific physical mechanism for envelope heating, e.g., from wave deposition in single massive stars  \citep{Piro11,Quataert12}, or from merger or common envelope evolution in massive-star binaries
\citep{Ivanova13,Podsiadlowski13,Justham14}.
An overall goal would be to assess the applicability of such super-Eddington winds for understanding inferred episodes of enhanced mass-loss from massive stars, for example LBV giant eruptions, or pre-SN events.

\section*{Acknowledgments}
The research here was initiated and substantially developed while the authors were participants in a massive-star research program at the Kavli Institute for Theoretical Physics, 
and as such was supported in part by the National Science Foundation under Grant No. NSF PHY-1125915. 
EQ was supported in part by a Simons Investigator award from the Simons Foundation and by the Gordon and Betty Moore Foundation through Grant GBMF5076.
RHDT acknowledges support from NASA grant NNX14AB55G and NSF grant ACI-1339606.
SPO was supported in part by NASA grant NNX15AM96G,
and also acknowledges sabbatical leave support from the University of Delaware, without which the collaborative research herein would likely not have occurred.
We thank Y.-F. Jiang for pointing out the importance of the radiation-advection term in the momentum and diffusion equations. 
We also thank K. Gayley, N. Shaviv and J. Vink for helpful comments on an early draft.
Finally, we thank the referee, Achim Feldmeier, for thoughtful comments and suggestions that helped improve the paper.

\bibliographystyle{mn2e}
\bibliography{OwockiS}

\begin{thebibliography}{}

\bibitem[\protect\citeauthoryear{{Begelman}}{{Begelman}}{2001}]{Begelman01}
{Begelman} M.~C.,  2001, \apj, 551, 897

\bibitem[\protect\citeauthoryear{{Begelman} \& {Rees}}{{Begelman} \&
  {Rees}}{1983}]{Begelman83}
{Begelman} M.~C.,  {Rees} M.~J.,  1983, in {Ferrari} A.,  {Pacholczyk} A.~G.,
  eds, Astrophysical Jets Vol.~103 of Astrophysics and Space Science Library,
  {Supercritical jets from a 'cauldron'}.
pp 215--225

\bibitem[\protect\citeauthoryear{{Chen}, {Woosley}, {Heger}, {Almgren} \&
  {Whalen}}{{Chen} et~al.}{2014}]{Chen14}
{Chen} K.-J.,  {Woosley} S.,  {Heger} A.,  {Almgren} A.,    {Whalen} D.~J.,
  2014, \apj, 792, 28

\bibitem[\protect\citeauthoryear{{Davidson} \& {Humphreys}}{{Davidson} \&
  {Humphreys}}{2012}]{Davidson12}
{Davidson} K.,  {Humphreys} R.~M.,  2012, \nat, 486, E1

\bibitem[\protect\citeauthoryear{{Foley}, {Smith}, {Ganeshalingam}, {Li},
  {Chornock} \& {Filippenko}}{{Foley} et~al.}{2007}]{Foley07}
{Foley} R.~J.,  {Smith} N.,  {Ganeshalingam} M.,  {Li} W.,  {Chornock} R.,
  {Filippenko} A.~V.,  2007, \apjl, 657, L105

\bibitem[\protect\citeauthoryear{{Fraser}, {Kotak}, {Pastorello}, {Jerkstrand},
  {Smartt}, {Chen}, {Childress}, {Gilmore}, {Inserra}, {Kankare}, {Margheim},
  {Mattila}, {Valenti}, {Ashall}, {Benetti}, {Botticella} \& {et al.}}{{Fraser}
  et~al.}{2015}]{Fraser15}
{Fraser} M.,  {Kotak} R.,  {Pastorello} A.,  {Jerkstrand} A.,  {Smartt} S.~J.,
  {Chen} T.-W.,  {Childress} M.,  {Gilmore} G.,  {Inserra} C.,  {Kankare} E.,
  {Margheim} S.,  {Mattila} S.,  {Valenti} S.,  {Ashall} C.,  {Benetti} S.,
  {Botticella} M.~T.,    {et al.} 2015, \mnras, 453, 3886

\bibitem[\protect\citeauthoryear{{Fraser}, {Magee}, {Kotak}, {Smartt}, {Smith},
  {Polshaw}, {Drake}, {Boles}, {Lee}, {Burgett}, {Chambers}, {Draper},
  {Flewelling}, {Hodapp} \& {et~al.}}{{Fraser} et~al.}{2013}]{Fraser13a}
{Fraser} M.,  {Magee} M.,  {Kotak} R.,  {Smartt} S.~J.,  {Smith} K.~W.,
  {Polshaw} J.,  {Drake} A.~J.,  {Boles} T.,  {Lee} C.-H.,  {Burgett} W.~S.,
  {Chambers} K.~C.,  {Draper} P.~W.,  {Flewelling} H.,  {Hodapp} K.~W.,
  {et~al.} 2013, \apjl, 779, L8

\bibitem[\protect\citeauthoryear{{Hindmarsh}}{{Hindmarsh}}{1983}]{Hindmarsh83}
{Hindmarsh} A.~C.,  1983, in {Stepleman} R.~S.,  ed., Scientific Computing
  Vol.~1 of IMACS Transactions on Scientific Computation, {ODEPACK: A
  Systematized Collection of ODE Solvers}.
pp 55--64

\bibitem[\protect\citeauthoryear{{Humphreys} \& {Davidson}}{{Humphreys} \&
  {Davidson}}{1994}]{Humphreys94}
{Humphreys} R.~M.,  {Davidson} K.,  1994, \pasp, 106, 1025

\bibitem[\protect\citeauthoryear{{Humphreys}, {Davidson}, {Jones}, {Pogge},
  {Grammer}, {Prieto} \& {Pritchard}}{{Humphreys} et~al.}{2012}]{Humphreys12}
{Humphreys} R.~M.,  {Davidson} K.,  {Jones} T.~J.,  {Pogge} R.~W.,  {Grammer}
  S.~H.,  {Prieto} J.~L.,    {Pritchard} T.~A.,  2012, \apj, 760, 93

\bibitem[\protect\citeauthoryear{{Ivanova}, {Justham}, {Chen}, {De Marco},
  {Fryer}, {Gaburov}, {Ge}, {Glebbeek}, {Han}, {Li}, {Lu}, {Marsh},
  {Podsiadlowski}, {Potter}, {Soker}, {Taam}, {Tauris}, {van den Heuvel} \&
  {Webbink}}{{Ivanova} et~al.}{2013}]{Ivanova13}
{Ivanova} N.,  {Justham} S.,  {Chen} X.,  {De Marco} O.,  {Fryer} C.~L.,
  {Gaburov} E.,  {Ge} H.,  {Glebbeek} E.,  {Han} Z.,  {Li} X.-D.,  {Lu} G.,
  {Marsh} T.,  {Podsiadlowski} P.,  {Potter} A.,  {Soker} N.,  {Taam} R.,
  {Tauris} T.~M.,  {van den Heuvel} E.~P.~J.,    {Webbink} R.~F.,  2013, \aapr,
  21, 59

\bibitem[\protect\citeauthoryear{{Jiang}, {Cantiello}, {Bildsten}, {Quataert}
  \& {Blaes}}{{Jiang} et~al.}{2015}]{Jiang15}
{Jiang} Y.-F.,  {Cantiello} M.,  {Bildsten} L.,  {Quataert} E.,    {Blaes} O.,
  2015, \apj, 813, 74

\bibitem[\protect\citeauthoryear{{Justham}, {Podsiadlowski} \&
  {Vink}}{{Justham} et~al.}{2014}]{Justham14}
{Justham} S.,  {Podsiadlowski} P.,    {Vink} J.~S.,  2014, \apj, 796, 121

\bibitem[\protect\citeauthoryear{{Kiewe}, {Gal-Yam}, {Arcavi}, {Leonard},
  {Emilio Enriquez}, {Cenko}, {Fox}, {Moon}, {Sand}, {Soderberg} \&
  {CCCP}}{{Kiewe} et~al.}{2012}]{Kiewe12}
{Kiewe} M.,  {Gal-Yam} A.,  {Arcavi} I.,  {Leonard} D.~C.,  {Emilio Enriquez}
  J.,  {Cenko} S.~B.,  {Fox} D.~B.,  {Moon} D.-S.,  {Sand} D.~J.,  {Soderberg}
  A.~M.,    {CCCP} T.,  2012, \apj, 744, 10

\bibitem[\protect\citeauthoryear{{Kochanek}}{{Kochanek}}{2011}]{Kochanek11}
{Kochanek} C.~S.,  2011, \apj, 743, 73

\bibitem[\protect\citeauthoryear{{Mauerhan}, {Smith}, {Filippenko},
  {Blanchard}, {Blanchard}, {Casper}, {Cenko}, {Clubb}, {Cohen}, {Fuller}, {Li}
  \& {Silverman}}{{Mauerhan} et~al.}{2013}]{Mauerhan13}
{Mauerhan} J.~C.,  {Smith} N.,  {Filippenko} A.~V.,  {Blanchard} K.~B.,
  {Blanchard} P.~K.,  {Casper} C.~F.~E.,  {Cenko} S.~B.,  {Clubb} K.~I.,
  {Cohen} D.~P.,  {Fuller} K.~L.,  {Li} G.~Z.,    {Silverman} J.~M.,  2013,
  \mnras, 430, 1801

\bibitem[\protect\citeauthoryear{{Meier}}{{Meier}}{1982}]{Meier82}
{Meier} D.~L.,  1982, \apj, 256, 706

\bibitem[\protect\citeauthoryear{{M{\"u}ller} \& {Vink}}{{M{\"u}ller} \&
  {Vink}}{2008}]{Mueller08}
{M{\"u}ller} P.~E.,  {Vink} J.~S.,  2008, \aap, 492, 493

\bibitem[\protect\citeauthoryear{{Nugis} \& {Lamers}}{{Nugis} \&
  {Lamers}}{2000}]{Nugis00}
{Nugis} T.,  {Lamers} H.~J.~G.~L.~M.,  2000, \aap, 360, 227

\bibitem[\protect\citeauthoryear{{Ofek}, {Sullivan}, {Cenko}, {Kasliwal},
  {Gal-Yam}, {Kulkarni}, {Arcavi}, {Bildsten}, {Bloom}, {Horesh}, {Howell},
  {Filippenko}, {Laher}, {Murray}, {Nakar}, {Nugent} \& {et al.}}{{Ofek}
  et~al.}{2013}]{Ofek13}
{Ofek} E.~O.,  {Sullivan} M.,  {Cenko} S.~B.,  {Kasliwal} M.~M.,  {Gal-Yam} A.,
   {Kulkarni} S.~R.,  {Arcavi} I.,  {Bildsten} L.,  {Bloom} J.~S.,  {Horesh}
  A.,  {Howell} D.~A.,  {Filippenko} A.~V.,  {Laher} R.,  {Murray} D.,  {Nakar}
  E.,  {Nugent} P.~E.,    {et al.} 2013, \nat, 494, 65

\bibitem[\protect\citeauthoryear{{Owocki} \& {Gayley}}{{Owocki} \&
  {Gayley}}{1997}]{Owocki97}
{Owocki} S.~P.,  {Gayley} K.~G.,  1997, in {Nota} A.,  {Lamers} H.,  eds,
  Luminous Blue Variables: Massive Stars in Transition Vol.~120 of Astronomical
  Society of the Pacific Conference Series, {ThePhysics of Stellar Winds Near
  the Eddingtson Limit}.
p.~121

\bibitem[\protect\citeauthoryear{{Owocki}, {Gayley} \& {Shaviv}}{{Owocki}
  et~al.}{2004}]{Owocki04}
{Owocki} S.~P.,  {Gayley} K.~G.,    {Shaviv} N.~J.,  2004, \apj, 616, 525

\bibitem[\protect\citeauthoryear{{Pastorello}, {Smartt}, {Mattila}, {Eldridge},
  {Young}, {Itagaki}, {Yamaoka}, {Navasardyan}, {Valenti}, {Patat}, {Agnoletto}
  \& {Augusteijn}}{{Pastorello} et~al.}{2007}]{Pastorello07}
{Pastorello} A.,  {Smartt} S.~J.,  {Mattila} S.,  {Eldridge} J.~J.,  {Young}
  D.,  {Itagaki} K.,  {Yamaoka} H.,  {Navasardyan} H.,  {Valenti} S.,  {Patat}
  F.,  {Agnoletto} I.,    {Augusteijn} T. e.~a.,  2007, \nat, 447, 829

\bibitem[\protect\citeauthoryear{{Piro}}{{Piro}}{2011}]{Piro11}
{Piro} A.~L.,  2011, \apjl, 738, L5

\bibitem[\protect\citeauthoryear{{Podsiadlowski}}{{Podsiadlowski}}{2013}]{Podsiadlowski13}
{Podsiadlowski} P.,  2013, in Massive Stars: From alpha to Omega {Evolved
  Binaries: Stellar Mergers, B[e] supergiants and LBVs}.
p.~130

\bibitem[\protect\citeauthoryear{{Press}, {Teukolsky}, {Vetterling} \&
  {Flannery}}{{Press} et~al.}{1992}]{Press92a}
{Press} W.~H.,  {Teukolsky} S.~A.,  {Vetterling} W.~T.,    {Flannery} B.~P.,
  1992, {Numerical recipes in FORTRAN. The art of scientific computing}

\bibitem[\protect\citeauthoryear{{Quataert}, {Fern{\'a}ndez}, {Kasen}, {Klion}
  \& {Paxton}}{{Quataert} et~al.}{2016}]{Quataert16}
{Quataert} E.,  {Fern{\'a}ndez} R.,  {Kasen} D.,  {Klion} H.,    {Paxton} B.,
  2016, \mnras, 458, 1214

\bibitem[\protect\citeauthoryear{{Quataert} \& {Shiode}}{{Quataert} \&
  {Shiode}}{2012}]{Quataert12}
{Quataert} E.,  {Shiode} J.,  2012, \mnras, 423, L92

\bibitem[\protect\citeauthoryear{{Shaviv}}{{Shaviv}}{1998}]{Shaviv98}
{Shaviv} N.~J.,  1998, \apjl, 494, L193

\bibitem[\protect\citeauthoryear{{Shen}, {Nakar} \& {Piran}}{{Shen}
  et~al.}{2016}]{Shen16}
{Shen} R.-F.,  {Nakar} E.,    {Piran} T.,  2016, \mnras, 459, 171

\bibitem[\protect\citeauthoryear{{Smith}}{{Smith}}{2014}]{Smith14}
{Smith} N.,  2014, \araa, 52, 487

\bibitem[\protect\citeauthoryear{{Smith}, {Gehrz}, {Hinz}, {Hoffmann}, {Hora},
  {Mamajek} \& {Meyer}}{{Smith} et~al.}{2003}]{Smith03b}
{Smith} N.,  {Gehrz} R.~D.,  {Hinz} P.~M.,  {Hoffmann} W.~F.,  {Hora} J.~L.,
  {Mamajek} E.~E.,    {Meyer} M.~R.,  2003, \aj, 125, 1458

\bibitem[\protect\citeauthoryear{{Smith} \& {Owocki}}{{Smith} \&
  {Owocki}}{2006}]{Smith06}
{Smith} N.,  {Owocki} S.~P.,  2006, \apjl, 645, L45

\bibitem[\protect\citeauthoryear{{van Marle}, {Owocki} \& {Shaviv}}{{van Marle}
  et~al.}{2009}]{vanMarle09}
{van Marle} A.~J.,  {Owocki} S.~P.,    {Shaviv} N.~J.,  2009, \mnras, 394, 595

\bibitem[\protect\citeauthoryear{{Vink}, {de Koter} \& {Lamers}}{{Vink}
  et~al.}{2000}]{Vink00}
{Vink} J.~S.,  {de Koter} A.,    {Lamers} H.~J.~G.~L.~M.,  2000, \aap, 362, 295

\end{thebibliography}

\appendix

\newpage
\section{Numerical Scheme} \label{sec:inoutsoln}

To solve the dimensionless momentum and diffusion equations
(\ref{eq:dwdx}) and (\ref{eq:dpdx}), we first transform $p$ into a new
variable $q \equiv p (1-x)^{-2}$, so that the equations become
\begin{align}
  \frac{{\rm d} w}{{\rm d}x} &= \Gamma_{\rm o} [1 - m(w + x) + 4 q \sqrt{w} ] - 1.
  \label{eq:A1}
  \\
  \frac{{\rm d} q}{{\rm d}x} &= - \frac{m \Gamma_{\rm o}}{\sqrt{w}} [ 1 - m(w + x)] + \frac{2 q}{1-x} + 4 m \Gamma_{\rm o} q, 
  \label{eq:A2}
\end{align}
(This transformation helps avoid numerical issues arising when $p$
becomes very small near the $x=1$ boundary). These equations, together
with the boundary conditions
\begin{equation}
  w(0) = 0, \qquad q(1) = 0
\end{equation}
comprise a two-point boundary value problem, which we solve using a
shooting technique \citep[see, e.g.,][]{Press92a}. Because the
boundaries are singular points, we integrate in both directions and
then stitch together the resulting solutions where they pass through
{\em radiative} sonic point $v=\sqrt{4/3} c_{\rm sr}$, such that
\begin{equation} \label{e:sonic}
w = \frac{4 q \sqrt{w}}{6 m}.
\end{equation}

For the outward integration, we choose an arbitrary initial
$q(0)=q_{0}$ at the inner boundary $x=0$,
and use a series expansion to write the solution at $x = \epsilon
\ll 1$ as
\begin{align}
 q(\epsilon) &= p_{0} - 2 m \Gamma_{\rm o} \sqrt{\frac{\epsilon}{\Gamma_{\rm o}-1}} + 2 p_{0} (1 + 2 m \Gamma_{\rm o}) \epsilon + \mathcal{O}(\epsilon^{3/2}), \\
 w(\epsilon) &= (\Gamma_{\rm o} - 1) \epsilon + \mathcal{O}(\epsilon^{2}).
\end{align}
Using these expressions as the starting point, we then integrate in
the direction of increasing $x$ using the \texttt{lsodar} routine from
the ODEPACK library of ordinary differential equation solvers
\citep{Hindmarsh83}, until one of three outcomes is realized:
\begin{enumerate}
\renewcommand{\theenumi}{(\roman{enumi})}
\item the integration is terminated before the $x = 1$ boundary is
  reached because $q < 0$.
\item the integration reaches the $x=1$ boundary with $q(1) > 0$.
\item the integration is terminated before the $x = 1$ boundary is
  reached because ${\rm d}q/{\rm d}x > 0$.
\end{enumerate}
Case (i) occurs when $q_{0}$ is chosen too small, while cases (ii) and
(iii) occur when $q_{0}$ is chosen too large. We apply a bisection
algorithm to determine the $q_{0}$ that yields case (ii) with the
smallest (positive) $q(1)$. 

In practice, the outward integration can be very sensitive to the chosen initial value $q_{0}$; 
indeed, for many combinations of $m$ and $\Gamma_{\rm o}$, an increase in just the 
last digit in the finite precision causes a switch from case (i) to case (iii), without 
a case (ii) that reaches $x=1$. 
When the integration does reach $x=1$,  the value $q_{0}$ is
insensitive to how small the numerically determined value of $q(1)$ is.
The bisection thus always tightly brackets the initial value $q_{0}$.
In any case this outward solution is only used to match a corresponding inward solution 
at the radiative sonic point, which is insensitive to the exact value $q(1)$.

For this inward integration, we choose an arbitrary initial $w_{1}$ at
the outer boundary $x=1$, and use a series expansion to write the
solution at $x = 1 - \epsilon$ as
\begin{align}
  w(1 - \epsilon) &= w_{1} + [1 + \Gamma_{\rm o} (m w_{1} + m - 1)] \epsilon + \mathcal{O}(\epsilon^{2}), \\
  q(1 - \epsilon) &= \frac{m \Gamma_{\rm o}}{3 \sqrt{w_{1}}} [1 - m (w_{1} + 1)] \epsilon + \mathcal{O}(\epsilon^{2})
\end{align}
We then use ODEPACK to integrate in the direction of decreasing $x$
until one of three outcomes is realized:
\begin{enumerate}
\renewcommand{\theenumi}{(\roman{enumi})}  
\item the integration is terminated before the $x = 0$ boundary is
  reached because $w < 0$.
\item the integration reaches the $x=0$ boundary with $w(0) > 0$.
\item the integration is terminated before the $x = 0$ boundary is
  reached, because ${\rm d}w/{\rm d}x < 0$.
\end{enumerate}
Case (i) occurs when $w_{1}$ is chosen too small, while cases (ii) and
(iii) occur when $w_{1}$ is chosen too large. We apply a bisection
algorithm to determine the $w_{1}$ which yields case (ii) with the
smallest (positive) $w(0)$.

\begin{figure}
\begin{center}
\includegraphics[scale=1.03]{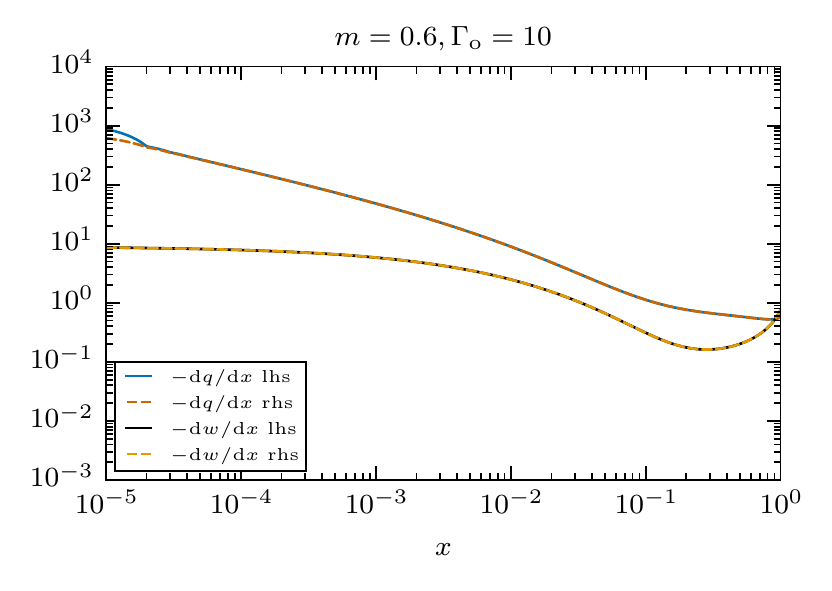}
\caption{
For a full numerical solution of the standard case with $\Gamma_{\rm o}=10$ and $m=0.6$,
comparison of spatial variations of the left-hand-sides (lhs; solid curves) and right-hand-sides (rhs; dashed curves) of the wind momentum equation 
(\ref{eq:A1}) (lower curves) and the transformed pressure equation (\ref{eq:A2}) (upper curves).
The fact that results for each equation both appear nearly as single curves illustrates  the nearly perfect overlap between the respective left and right sides, 
confirming that the derived numerical solution does indeed satisfy both of the original, coupled ODEs (\ref{eq:A1})  and (\ref{eq:A2}).}
\label{fig:residuals-mp6-G10}
\end{center}
\end{figure}

For the numerical solutions in the standard case with $\Gamma_{\rm o}=10$ and $m=0.6$, figure \ref{fig:residuals-mp6-G10}
shows that the numerically computed derivatives $dw/dx$ and $-dq/dx$ that appear on the left-hand-sides of the original coupled
ODEs (\ref{eq:A1}) and (\ref{eq:A2}) have a spatial variation that closely matches the terms that appear
on the corresponding right-hand-side.
This clearly demonstrates the accuracy of the derived numerical solution.

\section{Finite-sound-speed solutions}
\label{sec:finitecsg}

\begin{figure}
\begin{center}
\includegraphics[scale=0.65]{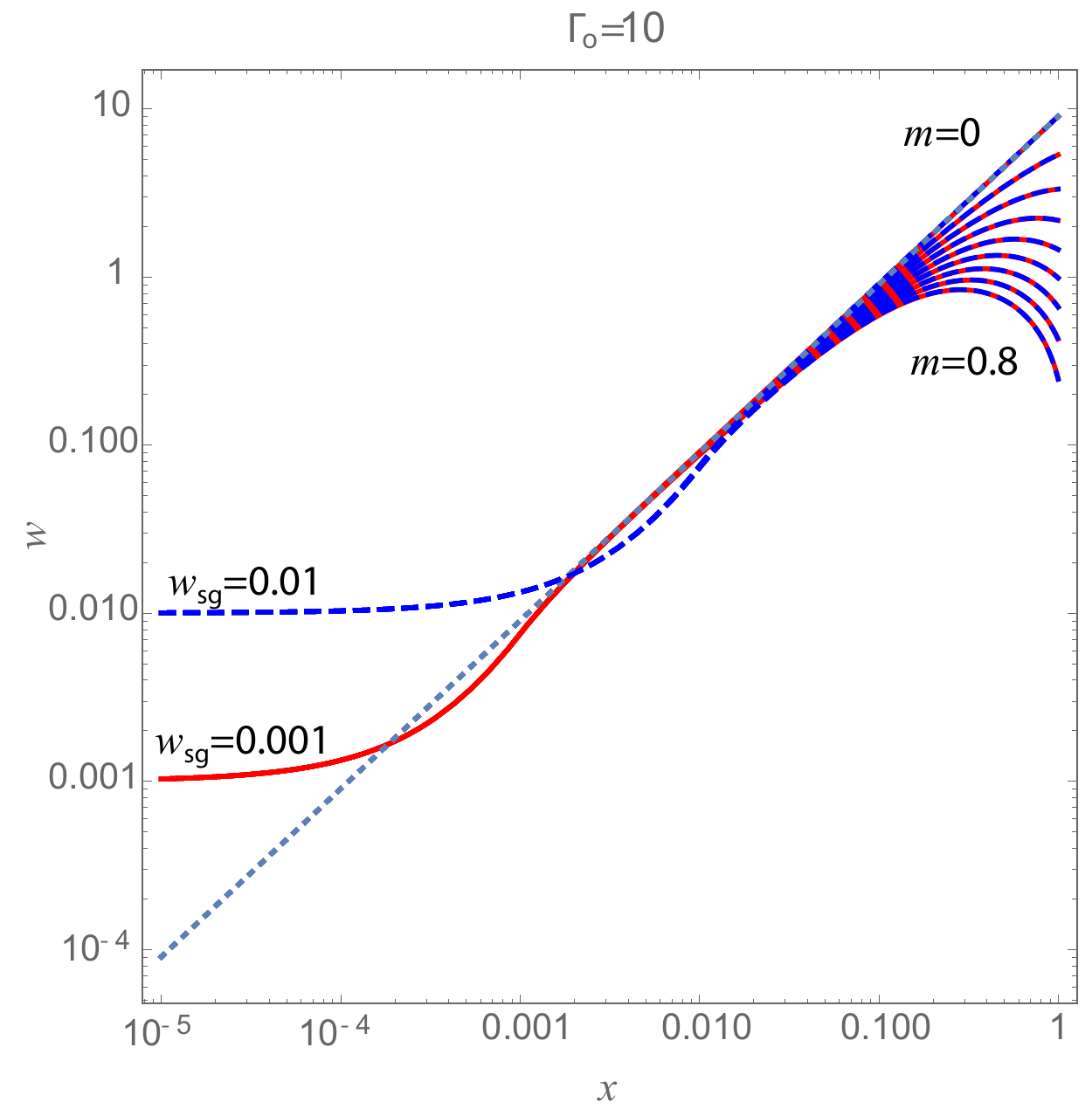}
\includegraphics[scale=0.65]{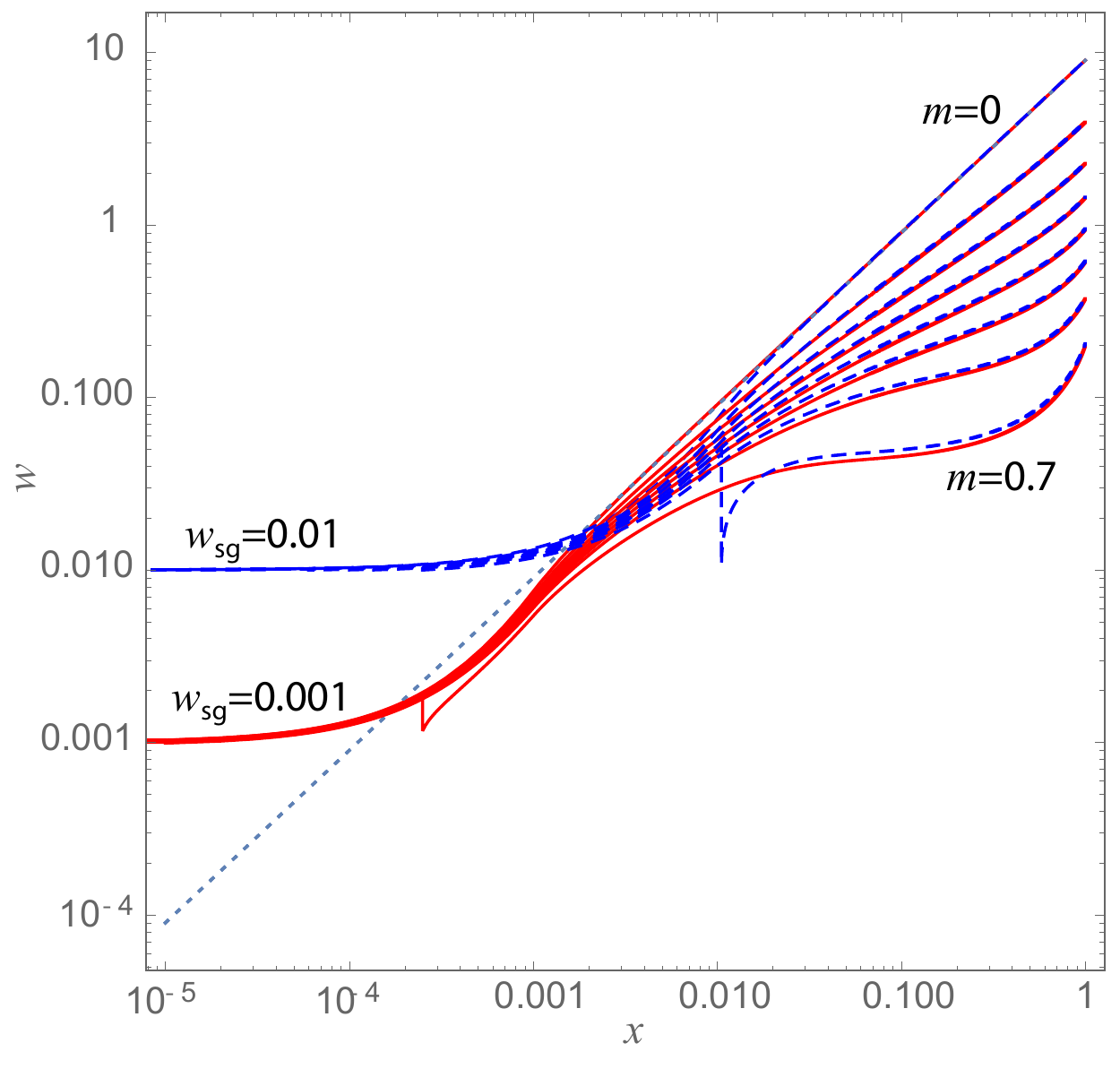}
\caption{
{\em Top:} For the case $\Gamma_{\rm o} = 10$ in the direct flux-driven model without radiative drag ($\eta=0$), scaled wind energy $w$ vs. $x$ for photon-tiring parameters $m=0$ to 0.8 in steps of 0.1, and for gas sound-speed cases $w_{\rm sg}=0.01$ (blue dashed curves) and $w_{\rm sg} = 0.001$ (red curves). The blue dotted line is the solution $w=(\Gamma_{\rm o}-1)x$ for the simple case with $m=w_{\rm sg}=0$.
{\em Bottom:} Same as top panel, but now for full solutions that account for radiation-pressure drag with a non-zero radiative enthalpy $\eta$.
The case $m=0.8$ no longer has a converged solution, and so is not shown.  The lowermost curve, for $m=0.7$, shows a discontinuity for both values of $w_{\rm sg}$, so is also not converged.
}
\label{fig:wxtirewsne0}
\end{center}
\end{figure}

To provide a smooth mapping onto a subsonic, nearly hydrostatic solution below the heating radius, let us now take a finite value for the gas sound speed on the left side of the momentum equation (\ref{eq:vdvdrcsg}).
The corresponding dimensionless momentum equation (\ref{eq:dwdxeta}) can then be recast in the form
\beq
\frac{dw}{dx} =  \frac{\Gamma (x)  ( 1 - m (w +x + \eta) )  -  1}{1 - {w_{\rm sg}}/{w}}
\, , 
\label{eq:dwdxws}
\eeq
where the model lower boundary $x=0$ is now taken at the gas sonic point, with $w(0)=w_{\rm sg}$.

We  also now assume that the heating is spread roughly over a scale height $H$, with an Eddington parameter that increases linearly  from a sonic point value $\Gamma_{\rm s}$ at $x=0$, to a final value $\Gamma_{\rm o}$ for $x \ge x_H$, 
\beq
\Gamma (x) = \Gamma_{\rm s} + (\Gamma_{\rm o} - \Gamma_{\rm s}) \min(x/x_H,1)
\, .
\label{eq:gamx}
\eeq
The sonic point value is set to ensure that the numerator of equation (\ref{eq:dwdxws}) vanishes,
\beq
\Gamma_{\rm s} \equiv \frac{1}{1-m(w_{\rm sg} + \eta_{\rm s})}
\, ,
\label{eq:gamsdef}
\eeq
while the spatial extent of the heating is given by
\beq
x_H \equiv 1 - \frac{R}{R+H} \approx 1- \left( 1 - \frac{H}{R} \right ) = \frac{H}{R}  \approx w_{\rm sg}
\, ,
\label{eq:xHdef}
\eeq
where the final approximation reflects the basic scaling of the scale height.

At the sonic point (where $w(x=0)=w_{\rm sg}$), the spatial gradient $w_{\rm s}'$ can be evaluated by applying L'Hopital's rule to the numerator and denominator in equation (\ref{eq:dwdxws}),
\beq
w_{\rm s}'^2 = (\Gamma_{\rm o}/\Gamma_{\rm s}  - 1)  -  m \Gamma_{\rm s} w_{\rm sg} (w_{\rm s}' +1 + \eta_{\rm s}' ))
\, .
\label{eq:wsLHopital}
\eeq
Since $w_{\rm s} \ll 1$, the term with a factor $m w_{\rm sg}$ is small compared to the order-unity first term in parentheses, implying that the sonic point slope is well approximated by
\beq
w_{\rm s}' \approx \sqrt{\Gamma_{\rm o}/\Gamma_{\rm s} -1 }
\, .
\label{eq:wsp}
\eeq

For the simple photon-tiring model that ignores radiation-pressure drag (so effectively sets $\eta=0$), the upper panel of figure \ref{fig:wxtirewsne0} presents a log-log plot of $w$ vs.\ $x$ for scaled sonic energies $w_{\rm sg} = 0.01$ (blue dashed curves) and $0.001$ (red curves), for photon tiring parameters ranging from $m=0$ to 0.8 in steps of 0.1.
The net result is to effectively truncate the wind energy to the finite, sonic-point value in the deep region $0<x<x_H$, but maintain the zero-sound-speed solution form in the outer wind. 

The lower panel of  figure \ref{fig:wxtirewsne0} shows analogous log-log plots for full solutions that include radiation-pressure drag and radiative enthalpy.
For both $w_{\rm sg}=10^{-3}$ and $10^{-2}$, the solutions for $m \le 0.6$ again reproduce the outer-wind results for the corresponding zero sound speed limit shown in the middle pane of figure \ref{fig:flow-vars-lin}, with just a shift over to the fixed sonic value in the inner wind.
However, for $m=0.7$,  this finite sonic energy implies also a strong radiation drag right from the sonic point, leading now to a mismatch between the outward and inward integrations discussed in Appendix \S A.
For still higher $m \ge 0.8$, this mismatch becomes even more pronounced,  so is not plotted.

The upshot is 
that including a small, but finite sound speed should have only minor effect on models with modest tiring parameters $m \lesssim 0.6$, but the coupling to the radiative drag can effectively preclude sonic point initiation of a wind for higher tiring parameters, $m \ge 0.7$.

\section{Role of convection}
\label{sec:convection}

We consider here the potential role of convection in carrying sufficient energy flux to keep the radiative luminosity below the Eddington limit. 
This can delay the initiation of a super-Eddington outflow to a higher, lower-density layer where convection becomes inefficient.
Writing
\beq
L_{\rm rad} = L_{\rm o} - L_{\rm c}
\, ,
\eeq
we can estimate the maximum convective flux as given by  free-streaming of the internal energy at some maximum convective speed $v_{\rm c,max}$,
\beq
\frac{L_{\rm c,max}}{ 4 \pi r^2} = v_{\rm c,max} E  \approx v_{\rm c,max} 3 P_{\rm rad} 
\, ,
\label{eq:Fconvmax}
\eeq
where the last equality assumes that, for the super-Eddington models here, the energy density is dominated by {\em radiation}.
Beyond this convective saturation, the energy flux must again be carried by radiation, leading to initiation of a super-Eddington outflow.
Writing this saturated convective luminosity as some factor $f$ of the Eddington value,  $L_{c,max} = f L_{\rm Edd}$,
the Eddington condition becomes $\Gamma_{\rm rad} = 1 = \Gamma_{\rm o} - f $ at some new wind initiation radius $r=R_{\rm o}$,
which solves to 
%\beqa
%\Gamma_{\rm o} - 1 = f &=& \frac{L_{\rm c,max}}{L_{\rm Edd}} 
%\nonumber
%\\
%&=&  \, \Gamma_{\rm o} \, \frac{4 \pi R_{\rm o}^2  \,  v_{\rm c,max} \, 3 P_{\rm rad,o} }{L_{\rm o}} 
%\nonumber
%\\
%&=&  3 \, \Gamma_{\rm o} p_{\rm o}  \, \frac{v_{\rm c,max}}{v_{\rm esc,o}}
%\, .
%\eeqa
\beq
\Gamma_{\rm o} - 1 = f = \frac{L_{\rm c,max}}{L_{\rm Edd}}  = 3 \, \Gamma_{\rm o} p_{\rm o}  \, \frac{v_{\rm c,max}}{v_{\rm esc,o}}
\, .
\eeq
Defining $\sqrt{w_{\rm cm}} \equiv v_{\rm c,max}/v_{\rm esc,o}$, the base value of the dimensionless radiation pressure is then
\beq
p_{\rm o} = \frac{\Gamma_{\rm o} -1}{3 \Gamma_{\rm o} \sqrt{w_{\rm cm}} }
\, .
\label{eq:po}
\eeq

Alternatively, using the fact that $m \eta = 4 p \sqrt{w}/(1-x)^2$,
% we can recast equation (\ref{eq:po}) in terms of the initial flow speed $v_{\rm o}$ at this radius for wind initiation; defining $\sqrt{w_{\rm o}} \equiv v_{\rm o}/v_{\rm esc,o}$, 
we find
\beq
m \eta_{\rm o} 
%=  \frac{4}{3} \, \frac{\Gamma_{\rm o} -1}{\Gamma_{\rm o} } \sqrt{\frac{w_{\rm o}}{w_{\rm cm}}}
= \frac{4}{3} \, \frac{\Gamma_{\rm o} -1}{\Gamma_{\rm o} } \frac{v_{\rm o}}{v_{\rm c,max}}
\, .
\eeq
Comparing with the dimensionless momentum equation (\ref{eq:dwdxeta}) at the lower boundary condition $w=x=0$, we see that the condition 
$dw/dx > 0$ for initiating an outflow becomes
\beq
 \frac{v_{\rm c,max}}{ v_{\rm o}} > \frac{4}{3}
\, .
\label{eq:gamolim}
\eeq

\begin{figure}
\begin{center}
\includegraphics[scale=0.65]{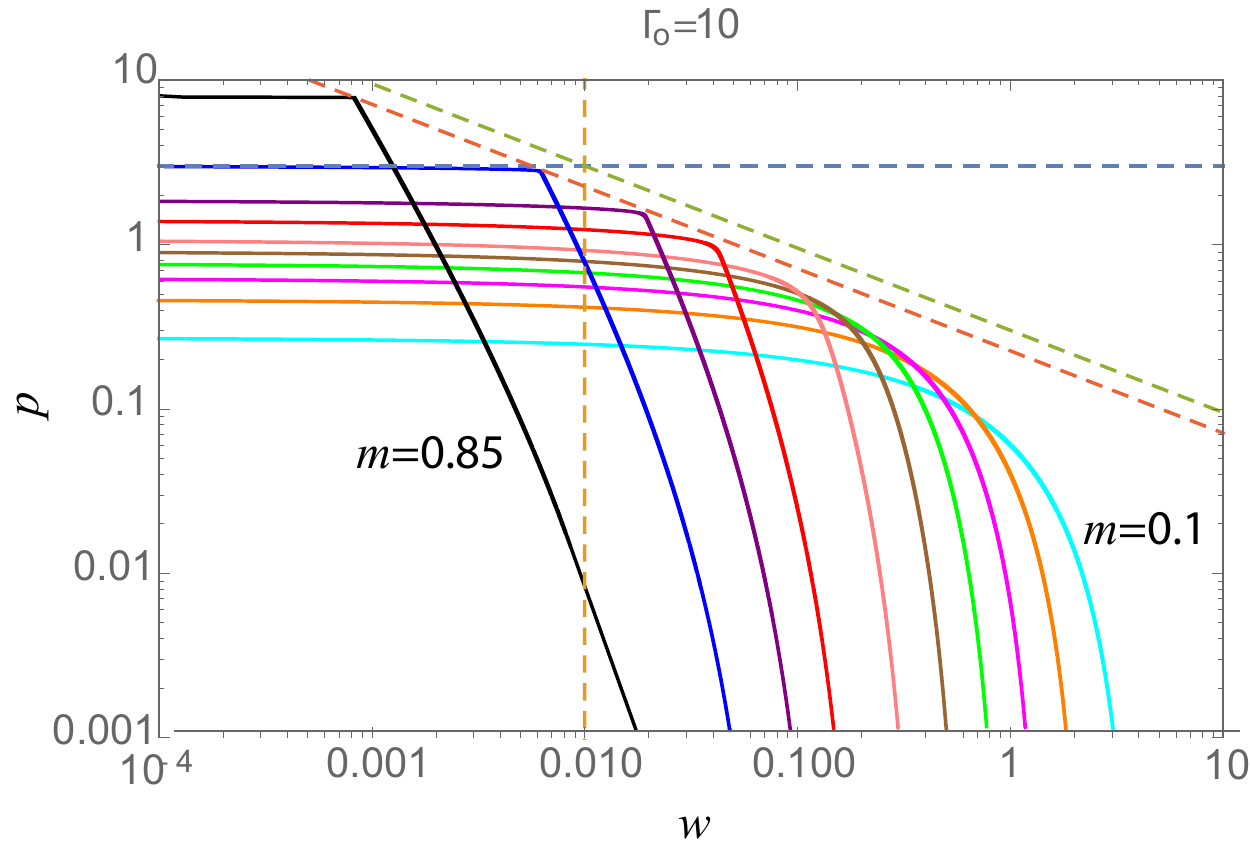}
\caption{For the zero-sound-speed, full-solutions with $\Gamma_{\rm o}=10$, plot of the scaled radiation pressure $p$ vs. the scaled wind energy $w$, for photon-tiring parameters
$m=0.1$ to 0.7 in steps of 0.1, plus $m= 0.75$, 0.8, and 0.85.
The dashed lines show various limit curves for the base pressure $p_{\rm o}$, for various assumptions of the sonic energy $w_{\rm sg}$ and the ratio of the maximum convective speed to sound speed, $v_{\rm c,max}/c_{\rm sg}$, as detailed in text.
}
\label{fig:pvswg10}
\end{center}
\end{figure}

Under the general assumption that the wind initial speed is given by the gas sound speed, $v_{\rm o}=c_{\rm sg}$, let us consider the consequences of the base pressure condition (\ref{eq:po}).
For the standard case of $\Gamma_{\rm o} = 10 $, figure \ref{fig:pvswg10} presents a log-log plot of $p$ vs. $w$ for $m=0.1$ to 0.7 in steps of 0.1, plus $m= 0.75$, 0.8, and 0.85, with black corresponding to $m=0.85$ near the photon-tiring limit, and cyan to weak photon-tiring $m=0.1$.

The vertical dashed gold line marks a canonical $w_{\rm sg}=0.01$, with the horizontal  dashed blue line the corresponding value of $p_{\rm o}$ from equations (\ref{eq:po}) if one takes the maximum convection speed to be given by this gas sound speed, i.e. $w_{\rm cm} = w_{\rm sg}$.  
The slanting green dashed line is the corresponding locus of $p_{\rm o}$ for variable $w_{\rm sg}$, still assuming $w_{\rm cm}=w_{\rm sg}$.

A key point is that these do not intersect the $p$ vs. $w$ curves for any $w_{\rm sg}$,  implying that an outflow initiated when convection is limited to $w_{\rm cm}=w_{\rm sg}$ cannot lead to a steady-state model.
This is consistent with the conclusion in equation (\ref{eq:gamolim}).

However,
 by equation (\ref{eq:gamolim}), 
even a small increase in maximum convection speed to $v_{\rm c,max}=(4/3) c_{\rm sg}$ allows wind initiation for any $\Gamma_{\rm o}$.
The slanting red dashed line shows the corresponding $p_{\rm o}$ variation for this case.  

Note that there are now intersections for the blue and black curves, corresponding to the $m=0.8$ and $m=0.85$ cases. 
But these require a small $w_{\rm sg}$, about 0.007 for $m=0.8$ (blue) and 0.001 for $m=0.85$ (black). 
Both these heavily tired models give very low terminal speed, viz. about $w(1) \approx 0.05$ for $m=0.8$, and $w(1) \approx 0.02$ for $m=0.85$.
Increasing $v_{\rm c,max}/c_{\rm sg}$ further would allow solutions for higher $w_{\rm sg}$, and at lower $m$ with higher $w(1)$.

Thus
delaying wind initiation to a point where convection becomes inefficient could lead to steady wind solutions, but these would tend to be near the photon-tiring limit $m \lesssim 1$, with very low terminal flow speed, $v_\infty/v_{\rm esc} = \sqrt{w(1)} \ll 1$.  
Moreover, the results depend quite sensitively on the details of the limiting convective speed $v_{\rm c,max}$, and the associated maximum convective energy flux.

\end{document}